\documentclass[%
 aip,
 amsmath,amssymb,
 reprint,%
]{revtex4-1}

\usepackage{graphicx}
\usepackage{dcolumn}
\usepackage{bm}
\usepackage{chemfig}
\usepackage{amsmath}
\usepackage{amssymb}

\usepackage[utf8]{inputenc}
\usepackage[T1]{fontenc}
\usepackage{mathptmx}
\usepackage{etoolbox}
\usepackage{float}
\usepackage{MnSymbol}
\usepackage{ulem}
\usepackage{xr}

\renewcommand{\t}{\text}
\newcommand{\Sec}[1]{Sec.\,\ref{#1}}
\newcommand{\Eq}[1]{Eq.\,(\ref{#1})}
\newcommand{\Fig}[1]{Fig.\,\ref{#1}}

\newcommand{\RNum}[1]{\uppercase\expandafter{\romannumeral #1\relax}}
\usepackage{tikz}

\newcommand{\au}{\,a.u.\,}

\makeatletter
\def\@email#1#2{%
 \endgroup
 \patchcmd{\titleblock@produce}
  {\frontmatter@RRAPformat}
  {\frontmatter@RRAPformat{\produce@RRAP{*#1\href{mailto:#2}{#2}}}\frontmatter@RRAPformat}
  {}{}
}%
\makeatother
\begin{document}

\preprint{AIP/123-QED}

\title{Harnessing multi-mode optical structure for chemical reactivity}
\author{Yaling Ke}
\email{yaling.ke@phys.chem.ethz.ch}
\author{Jakob Assan}
\affiliation{ 
Department of Chemistry and Applied Biosciences, ETH Zürich, 8093 Zürich, Switzerland
}%

\begin{abstract}
The prospect of controlling chemical reactivity using frequency-tunable optical microcavities has materialized over the past decade, evolving into a fascinating yet challenging new field of polaritonic chemistry, a multidisciplinary domain at the intersection of quantum optics, chemical dynamics, and non-equilibrium many-body physics. While most theoretical efforts to date have focused on single-mode cavities, practical implementations in polaritonic chemistry typically involve planar optical cavities that support a series of equally spaced photon modes, determined by the cavity geometry. In this work, we present a numerically exact, fully quantum-mechanical study of chemical reactions in few-mode cavities, revealing two key scenarios by which multi-mode effects can enhance cavity-modified reactivity. The first scenario emerges when the free spectral range is comparable to the single-mode Rabi splitting. In such cases, hybridization between a rate-decisive molecular vibration and a central resonant cavity mode reshapes the resonance landscape, enabling additional reaction pathways mediated by adjacent cavity modes. The second scenario exploits the intrinsic anharmonicity of molecular vibrations, which gives rise to multiple dipole-allowed transitions with distinct energies. Under multi-mode strong coupling, where different cavity modes individually resonate with these distinct transitions, multi-photon processes involving sequential absorption across multiple modes become accessible. This leads to a nontrivial and non-additive rate enhancement via cascade-like vibrational ladder climbing. Together, these findings offer new strategies for tailoring chemical reactivity by harnessing the structural richness of multi-mode structure, offering valuable insights for optimal experimental designs in polaritonic catalysis.
\end{abstract}

\maketitle

\section{Introduction}
Strong coupling between molecular vibrations and confined light fields--typically achieved using optical microcavities--has increasingly come into focus over the past decade.\cite{Ebbesen_2016_ACR_p2403} In particular, there is growing interest in understanding how light-matter interactions affect chemical reactivity, especially when cavity modes are tuned into resonance with specific vibrational absorption bands.\cite{Thomas_2016_ACE_p11634,Vergauwe_2019_ACIE_p15324,Lather_2019_ACIE_p10635,Thomas_2019_S_p615,Hiura_2019__p,Hirai_2020_ACE_p5370,Pang_2020_ACIE_p10436,Sau_2021_ACIE_p5712,Ahn_Sci_2023_p1165,Patrahau_Angew.Chem.Int.Ed._2024_p202401368} However, typical experimental scenarios involve a macroscopic ensemble of reactive molecules embedded in a condensed-phase solvent and collectively coupled to a delocalized electromagnetic field confined inside the cavity, which is often exposed to a lossy radiative environment. This configuration has qualified to become a complex system, potentially exhibiting pronounced sensitivity to subtle changes among its numerous interacting components. Moreover, interpreting the observed cavity-induced modifications in reaction rates arguably requires quantum-level descriptions\cite{Hu_2023_JPCL_p11208,Ke_2024_JCP_p54104} under non-equilibrium conditions,\cite{Li_2020_JCP_p234107,Lather_2022_CS_p195,ke2025non} adding yet another layer of complexity to the problem. It is fair to say that a rigorous study of chemical reactions inside an optical cavity now stands at the confluence of quantum optics, organic chemistry, complex system theory, and non-equilibrium many-body physics--posing formidable obstacles, yet offering exciting opportunities for discovery. 

Thus far, the majority of theoretical efforts have concentrated on the single-mode limit,\cite{Galego_2016_NC_p13841,Galego_2017_PRL_p136001,Galego_2019_PRX_p21057,CamposGonzalezAngulo_2019_NC_p4685,CamposGonzalezAngulo_2023_JCP_p230901,Mandal_2020_JPCL_p9215,Mandal_2022_JCP_p,Mandal_2023_CR_p9786,Yang_2021_JPCL_p9531,Li_2021_JPCL_p6974,Li_2021_NC_p1315,Li_2021_AC_p15661,Sun_2022_JPCL_p4441,Schaefer_2022_NC_p7817,Wang_2022_JPCL_p3317,Fischer_2022_JCP_p154305,Fiechter_2023_JPCL_p8261,Sokolovskii_2023_apa_p,Pavosevic_2023_NC_p2766,Ruggenthaler_2023_CR_p11191,Vega__2024_p,Lindoy_2023_NC_p2733,Lindoy_2024_N_p2617,Ying_2023_JCP_p84104,Ying_2024_CM_p110,Ke_J.Chem.Phys._2024_p224704,Ke_J.Chem.Phys._2025_p64702} assuming that only the resonant coupling to one confined mode of the photonic structure is at play, while interactions with all other photonic modes are either neglected or bulked into a continuum of photonic background.\cite{Carmichael_1989_PRA_p5516} The simplification is often justified in systems such as nanometer-scale cavities or surface plasmon resonators, where the free spectral range (FSR)--the energy spacing between adjacent cavity modes--is sufficiently large that only a single cavity mode falls within resonance with the molecular excitation of interest.\cite{Fischer_2002_PRL_p163002,Connolly_2003_APL_p5377,Brennecke_2007_N_p268,Terraciano_2009_NP_p480} In such cases, far-detuned cavity modes can be safely ignored. However, in most polaritonic chemistry experiments, particularly those involving solvent-cooperative vibrational strong coupling,\cite{Vergauwe_2019_ACIE_p15324,Chen_2024_N_p2591,Lather_2019_ACIE_p10635} the frequently used Fabry-P\'erot cavities are constructed from two parallel reflective mirrors separated by a few to tens of micrometers. These cavity geometries support a set of cavity modes that are more closely spaced in frequency. In some instances, the FSR becomes even comparable to the Rabi splitting--the energy gap between the two bright light-matter hybrid polaritonic states--making it no longer appropriate to ignore contributions from additional modes.\cite{George_2016_PRL_p153601,KenaCohen_2013_AOM_p827,Sundaresan_2015_PRX_p21035} Only recently have a handful of studies begun to explore the influence of this complex multi-mode structure on molecular observables,\cite{Coles_2014_APL_p191108,Hoffmann_2020_JCP_p104103, Georgiou_2021_JCP_p124309,Thomas_2024_JCP_p204303, Herrera_2024_PTA_p20230343,Ribeiro_2022_CC_p48,Engelhardt_2023_PRL_p213602,Mandal_2023_NL_p4082} such as vibrational absorption and emission properties.\cite{Simpkins_2015_AP_p1460,Tichauer_2021_JCP_p104112,Godsi_2023_JCP_p134307,Fischer_2022_JCP_p34305,Menghrajani_2024_JPCL_p7449} Nevertheless, the full implications of multimode coupling for chemical reactivity remain largely unexplored.

In this work, we present a fully quantum dynamical investigation of chemical reactions in a few-mode optical cavity, aiming to demonstrate how multi-mode effects can be harnessed to optimize cavity-enhanced catalysis. We consider a prototypical molecular system represented by a double-well potential, as established in previous studies,\cite{Lindoy_2023_NC_p2733,Lindoy_2024_N_p2617,Ying_2023_JCP_p84104,Hu_2023_JPCL_p11208,Ying_2024_CM_p110,Ke_J.Chem.Phys._2024_p224704,Ke_2024_JCP_p54104,Ke_J.Chem.Phys._2025_p64702} and simulate its reactive dynamics using the numerically exact hierarchical equations of motion (HEOM) approach in conjunction with a tree tensor network state (TTNS) solver.\cite{Ke_2023_JCP_p211102}  We explore two key scenarios in which additional discrete cavity modes lead to further enhancement of reaction rates inside the cavity.  In the first scenario, we consider a low-finesse cavity where supported cavity modes are spectrally separated by a relatively small FSR. In this case, multiple cavity modes may lie close in energy. Tuning a high-order cavity mode into resonance with a molecular vibrational transition allows neighboring cavity modes to also couple effectively with the resulting hybrid polaritonic states, which retain partial molecular characters. This collective resonance opens new pathways for reactivity, thereby enhancing the overall reaction rates. Notably, this additional catalytic effect in the presence of the neighboring modes is most pronounced when the FSR is comparable to single-mode Rabi splitting. In the second scenario, we explore how the intrinsic anharmonicity of molecular vibrations--characterized by multiple vibrational transitions at distinct energies--can be exploited under multi-mode strong coupling. By tailoring different cavity modes to individually match distinct vibrational transitions, a spontaneous multi-photon process is facilitated via a sequential cascade of vibrational transitions. This process enables a non-additive, synergistic rate enhancement, which cannot be explained by considering each mode in isolation. In addition, we highlight the connection and critical distinction between molecular absorption spectra and the cavity-modified reaction rate profile plotted as a function of the cavity frequency. In line with our previous work,\cite{ke2025non} our results underscore that reaction dynamics are governed by more than just the thermal population of vibrational levels and the strength of their transition dipoles--factors that shape the linear absorption spectra. More importantly, the dynamical interplay among multiple vibrational pathways and different energy exchange processes, modulated by the cavity field, also plays a decisive role.

The remainder of the paper is organized as follows. In \Sec{sec:theory}, we introduce the microscopic model system and the theoretical approaches used to simulate chemical reactions in multi-mode optical microcavities. \Sec{sec:results} describes the numerical details and presents the numerical results, illustrating the two aforementioned scenarios in which multiple cavity modes have nontrivial impacts on reaction rates. Finally,  \Sec{sec:conclusion} summarizes our findings and discusses potential extensions of multi-mode cavity-modified chemistry in future work.
 
\begin{figure}\centering
  \begin{minipage}[c]{0.45\textwidth}
    \raggedright a) 
    \includegraphics[width=\textwidth]{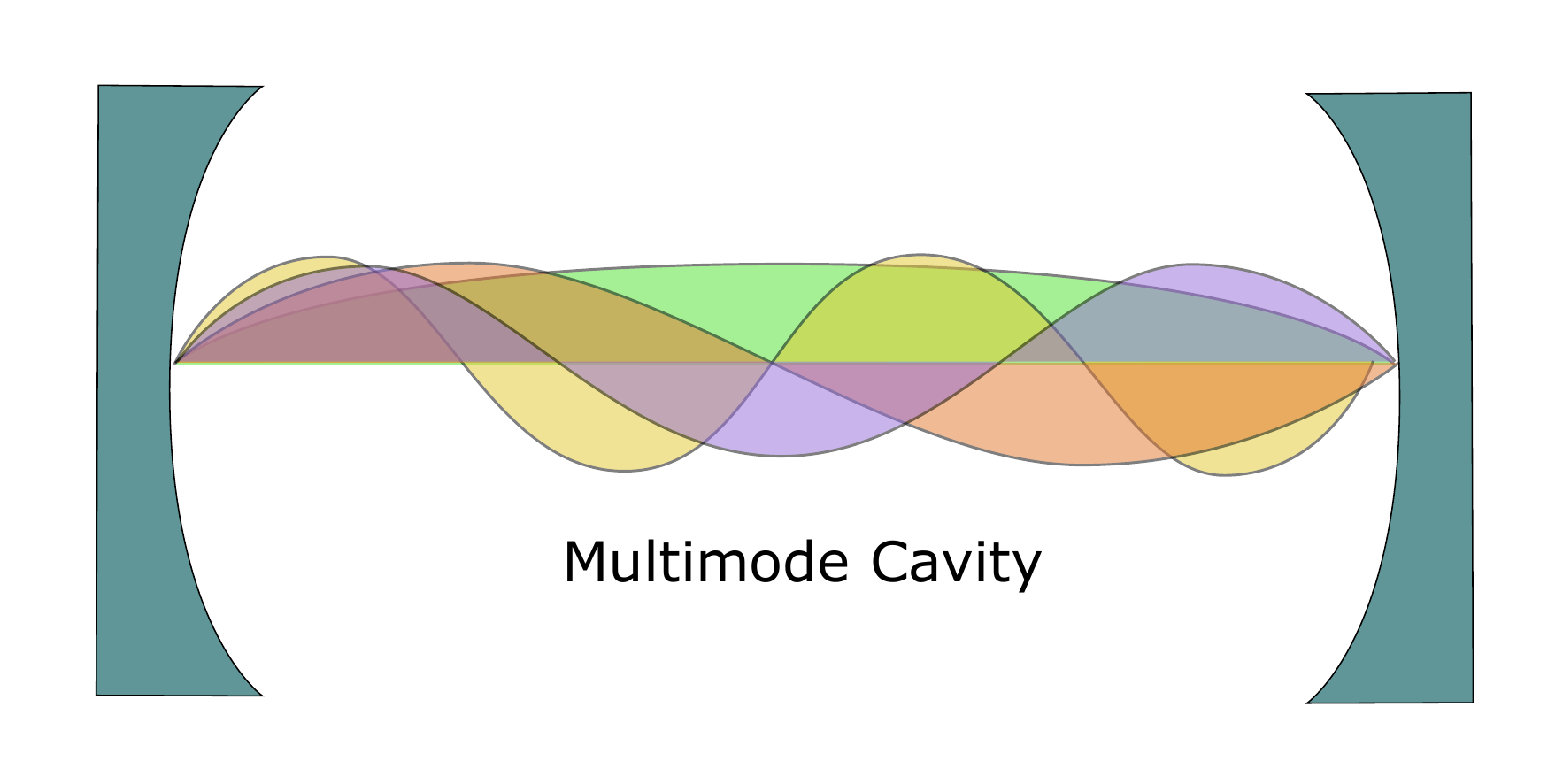}\\
    \raggedright b) 
  \end{minipage}
  \begin{minipage}[c]{0.35\textwidth}
    \includegraphics[width=\textwidth]{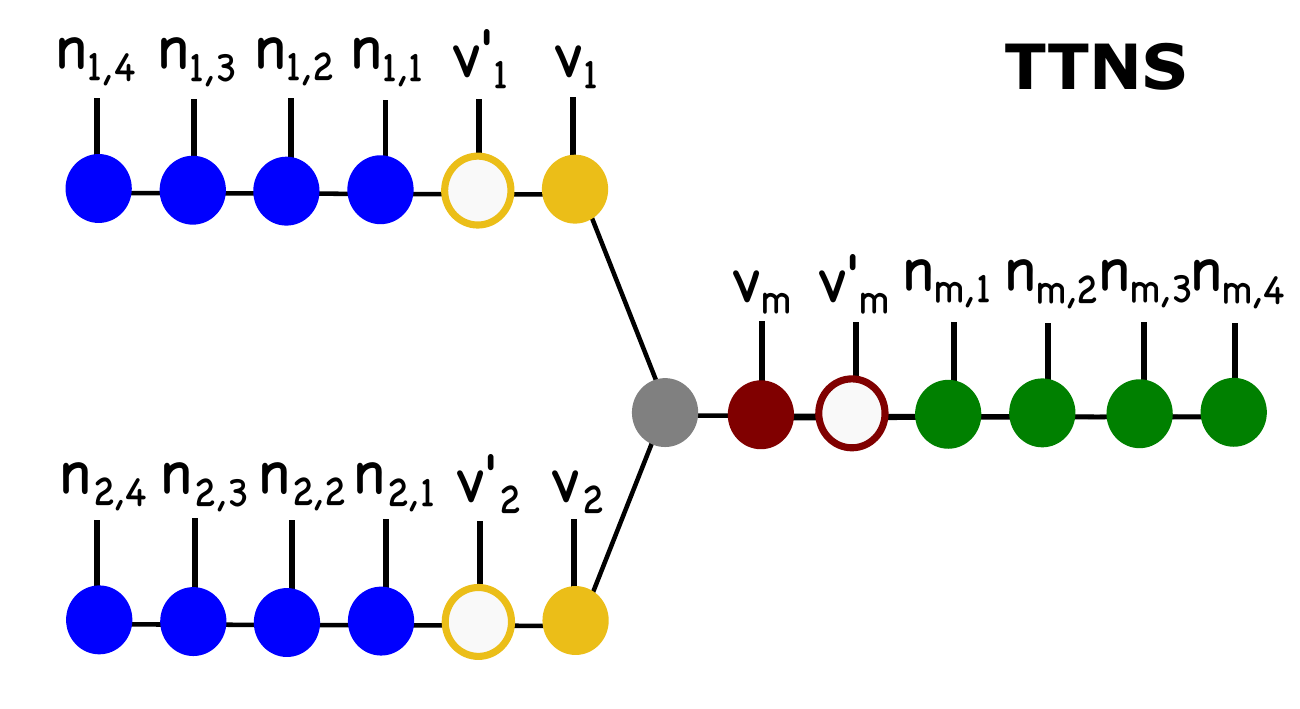}
  \end{minipage}
\caption{a) Sketch of an optical cavity with a path length of $L$ that supports multiple discrete cavity modes. b) Graphical representation of the TTNS decomposition for the extended wavefunction $|\Psi(t)\rangle$ for an open quantum system model describing chemical reactions in optical cavity within the HEOM framework. Each colored node with an open leg in the diagram represents a low-rank tensor related to a physical DoF (red: molecule, yellow: two cavity modes, green: solvent, and blue: cavity baths). Gray node is a rank-3 connecting tensor, introduced for improving the numerical efficiency.  Here, both the solvent and cavity baths are represented by four Pad\'e poles, i.e., $P=4$ in \Eq{timecorrelation}. A connected leg represents a shared virtual index between two nodes, which runs from $1$ to $D_{i}$. The maximal bond dimension is denoted as $D_{\rm max}$.} \label{fig1:schematicplot}
\end{figure}

\section{\label{sec:theory}Theory}
In this work, we consider a single molecule placed inside a planar microcavity that supports multiple confined radiation modes. The molecular system within this multimode cavity is described using the Pauli-Fierz light-matter Hamiltonian in the dipole gauge under the long-wavelength approximation,\cite{Flick_2017_PotNAoS_p3026,Rokaj_2018_JPBAMOP_p34005,Mandal_2023_CR_p9786,Lindoy_2023_NC_p2733} with $\hbar=1$ employed throughout:
\begin{equation}
\label{Hs}
\begin{split}
    H_{\rm S} =  \frac{p_{\rm m}^2}{2M} + U(x_{\rm m})  +
    \sum_{i}\frac{p_{i}^2}{2} + \frac{1}{2} \omega_{i}^2\left(x_{i} + \sqrt{\frac{2}{\omega_{i}}} \eta_{i} \vec{\mu}(x_{\rm m})\cdot \vec{e}_i \right)^2.
\end{split}
\end{equation}
Here, $x_{\rm m}$ and $p_{\rm m}$ represent the reactive coordinate and its conjugated momentum, respectively, with $M$ denoting the reduced mass associated with the reactive bond. In addition, we focus exclusively on ground-state reactions, with the corresponding electronic potential energy surface denoted by $U(x_{\rm m})$. 

The cavity has an optical path length $L$, and the mode dispersion is given by $\omega_i=\frac{c}{n_d}\sqrt{{q_{\perp}^2(i)}+q_{\parallel}^2}$, where $c$ is the speed of light, $n_d$ is the dielectric constant of the intracavity medium. The transverse wavevector component  $q_{\perp}(i) = \frac{i\pi}{2L}$ is perpendicular to the mirrors and discretized by boundary conditions, allowing only non-zero positive integers $i$, while $q_{\parallel}$ represents the quasi-continuous in-plane wavevector. At the normal incidence ($q_{\parallel}=0$), which is assumed throughout this work, the cavity supports a series of standing-wave modes that are uniformly spaced in frequency, as schematically illustrated in \Fig{fig1:schematicplot}~a). The spacing, known as the free spectral range, is given by $\Delta = \frac{\pi c}{2n_d L}$, corresponding to the frequency separation between adjacent cavity modes. For low-finesse cavities or those with an extended path length, the FSR can become comparable to or even smaller than the Rabi splitting,\cite{Balasubrahmaniyam_2021_PRB_p241407} which is the energy separation between the bright upper and lower polaritons formed through the light-matter hybridization. The dispersionless cavity mode at the $i$th mode number $\omega_i=i\Delta$ is modeled as a quantum harmonic oscillator with coordinate $q_{i}$, conjugate momentum $p_{i}$, and frequency $\omega_i$. The cavity oscillators are displaced due to light-matter interaction, characterized by a coupling strength $\eta_{i}$ for each mode $i$. The unit vector $\vec{e}_i$ specifies the light polarization, while the molecular transition dipole moment, $\vec{\mu}(x_{\rm m})$, which depends on the reaction coordinate $x_{\rm m}$, mediates the interaction between the molecule and the confined cavity field. Here, we have focused on the $q_{\parallel}=0$ limit, while dispersion effects at finite $q_{\parallel}$ remain an open direction for future investigation.

To account for dissipation, we consider two types of bosonic environments. First, we assume that the non-reactive molecular vibrational modes, as well as the surrounding solvent, are infrared-inactive and collectively form a bosonic reservoir composed of an infinite set of harmonic oscillators, referred to as the molecular bath:
\begin{equation}
\label{environmentHamiltonian}
\begin{split}
H_{\rm E}^{\rm m} = &
 \sum_{k} \frac{P^2_{{\rm m} k}}{2}+\frac{1}{2}\omega_{{\rm m}k}^2 \left(Q_{{\rm m}k}+\frac{g_{{\rm m}k}x_{\rm m}}{\omega_{{\rm m}k}^2}\right)^2 
\end{split}
\end{equation}
Each bath oscillator, indexed by $k$, is characterized by its coordinate $Q_{{\rm m} k}$,  conjugate momentum $P_{{\rm m} k}$, and frequency $\omega_{{\rm m} k}$. It couples to the reactive coordinate with a coupling strength $g_{\rm m k}$, resulting in a displacement of $\frac{g_{{\rm m}k}x_{\rm m}}{\omega_{{\rm m}k}^2}$. 

Second, to capture cavity losses, we introduce a separate bosonic bath for each cavity mode. These baths represent the continuum of external electromagnetic modes that interact with the confined photonic modes. The Hamiltonian for the cavity bath interacting with the $i$th cavity mode is given by
\begin{equation}
\label{environmentHamiltonian}
H_{\rm E}^{\rm i} = \sum_{k} \frac{P_{{\rm i}k}^2}{2}+ \frac{1}{2}\omega_{{\rm i}k}^2 \left(Q_{{\rm i}k}+\sum_i\frac{g_{ik}}{\omega_{{\rm i}k}^2}x_{i}\right)^2,
\end{equation}
where $Q_{ik}$, $P_{ik}$, $\omega_{ik}$, and $g_{ik}$ characterize the coordinate, momentum, frequency, and coupling strength of the $k$th oscillator in the photonic bath for photonic mode $i$. We also explore the scenario where all cavity modes are coupled to a single shared cavity bath; the corresponding results are presented in the Supplementary Information (SI).

The total Hamiltonian is thus given by $H=H_{\rm S}+H^{\rm m}_E+\sum_iH^{\rm i}_E$, and the whole coupled system is treated within the framework of open quantum dynamics.\cite{Lindoy_2023_NC_p2733,Lindoy_2024_N_p2617,Fiechter_2023_JPCL_p8261,Ying_2023_JCP_p84104,Hu_2023_JPCL_p11208,Ying_2024_CM_p110,Ke_J.Chem.Phys._2024_p224704,Ke_2024_JCP_p54104,Ke_J.Chem.Phys._2025_p64702} 
The quantum dynamics and equilibrium properties of the coupled molecule-cavity system in the dissipative environment can be obtained within the HEOM framework (see Ref.~\onlinecite{Tanimura_2020_JCP_p20901} and the references therein), which is a numerically exact open quantum system approach. The method leverages the exponential expansion of the two-time correlation function of a Gaussian bath, which reads 
\begin{equation}
\label{timecorrelation}
    C_{\alpha}(t) = \frac{1}{\pi} \int_{-\infty}^{\infty} \frac{e^{-i\omega t}}{1-e^{-\beta \omega}} J_{\alpha}(\omega) \mathrm{d}\omega = \sum_{p=1}^{P\rightarrow \infty} \lambda_{\alpha}^2\eta_{\alpha p} e^{-i\gamma_{\alpha p} t}.
\end{equation}
Here, the spectral density function $J_{\alpha}(\omega)=\frac{\pi}{2}\sum_k \frac{g_{\alpha k}^2}{\omega_{\alpha k}}\delta(\omega -\omega_{\alpha k})$ characterizes the energetic distribution of the molecular bath $(\alpha={\rm m})$ or the cavity bath for the $i$th cavity mode $(\alpha=i)$.  The reorganization energy $\lambda_{\alpha}^2=\frac{1}{\pi}\int_0^{\infty}\frac{J_{\alpha}(\omega)}{\omega}\mathrm{d}\omega$ quantifies the overall coupling strength between the system DoF $\alpha$ and its own bath. Each term in the exponential expansion in \Eq{timecorrelation} represents an effective bosonic pseudomode, with the exponent $\gamma_{\alpha p}$ interpreted as the effective frequency and the prefactor $\eta_{\alpha p}$ as the corresponding coupling strength. Truncating the expansion at a suitable number of terms $P$ where the desired accuracy is achieved, either through the analytical expressions or numerical fitting procedures,\cite{Hu_2010_JCP_p101106,Xu_2022_PRL_p230601}  provides a feasible and accurate means of modeling the original continuum bath using only a finite set of effective modes, while preserving all essential statistical information.

Furthermore, by introducing second quantization for these discrete dissipative pseudomodes with the Fock state $|\mathbf{n}\rangle=|\cdots n_{\alpha p}\cdots \rangle$ in the number representation--where $n_{\alpha p}$ indexes over the non-negative integers--and the associated bosonic creation and annihilation operators, $b_{\alpha p}^{+}$ and $b_{\alpha p}$, defined as
\begin{subequations}
\begin{equation}
      b_{\alpha p}^{+} |\mathbf{n}\rangle = \sqrt{n_{\alpha p}+1}| \mathbf{n}_{\alpha p}^+ \rangle;
\end{equation}  
\begin{equation}
      b_{\alpha p} |\mathbf{n}\rangle = \sqrt{n_{\alpha p}}|\mathbf{n}_{\alpha p}^- \rangle,
\end{equation}
\end{subequations}
where $|\mathbf{n}_{\alpha p}^{\pm}\rangle=|\cdots, n_{\alpha p}\pm 1, \cdots\rangle$, the HEOM can be formulated as a time-dependent Schr\"odinger-like equation:\cite{Borrelli_2019_JCP_p234102,Borrelli_2021_WCMS_p1539,Ke_2022_JCP_p194102} 
\begin{equation}
\label{Schroedinger}
    \frac{\mathrm{d}|\Psi(t)\rangle}{\mathrm{d} t} = -i\mathcal{H} |\Psi(t)\rangle
\end{equation}
for the extended wave function 
\begin{equation}
|\Psi(t)\rangle=\sum_{\bf{v,n}}C_{\bf{v,n}}|v^{}_{m}v'_{m}\rangle\otimes |v^{}_1v'_1\rangle \otimes \cdots\otimes |v^{}_iv'_i\rangle\otimes \cdots\otimes|n_{\alpha p}\rangle\otimes\cdots,
\end{equation}
where the indices $v_{m/i}$ and $v'_{m/i}$ refer to bra and ket components of the system DoFs in the density matrix. The non-Hermitian super-Hamiltonian $\mathcal{H}$ is explicitly given by 
\begin{equation}
\begin{split}
    \mathcal{H} = &\hat{H}_{\rm S}+\sum_{\alpha}\lambda_{\alpha}^2 \hat{q}_{\alpha}^2 -\tilde{H}_{\rm S}-\sum_{\alpha}\lambda_{\alpha}^2 \tilde{q}_{\alpha}^2 -i \sum_{\alpha}\sum_p  \gamma_{\alpha p} b_{\alpha p}^{+}b_{\alpha p}\\
    & +\sum_{\alpha}\sum_p \lambda_{\alpha}\left[(\hat{q}_{\alpha}-\tilde{q}_{\alpha})b_{\alpha p}     +(\eta_{\alpha p}\hat{q}_{\alpha}-\eta^*_{\alpha p}\tilde{q}_{\alpha})b_{\alpha p}^+ \right].
\end{split}
\end{equation}
Each system operator $O_{j}$ for the $j$th system DoF in Hilbert space is associated with a pair of superoperators in twin space: $\hat{O}_{j}=O_{j}\otimes I_{j}$ and $\tilde{O}_{j}=I_{j}\otimes O_{j}^{\dagger}$, where $I_{j}$ denotes the identity operator for the $j$th DoF.  To efficiently simulate the time-dependent wavefunction, we decompose the high-rank coefficient tensor $C_{\bf{v,n}}$ in the extended wave function $|\Psi(t)\rangle$ using a binary tree tensor network state, as exemplarily illustrated in \Fig{fig1:schematicplot}~b) for the case of a two-mode cavity, which is optimal for the star-like topology where the molecule is coupled to many cavity and bath modes. Implementation details can be found in Refs.\,\onlinecite{Ke_J.Chem.Phys._2025_p64702} and \onlinecite{Ke_2023_JCP_p211102}.

\section{\label{sec:results}Results}
As a first step toward a broader understanding of how multi-mode cavities influence chemical reactivity, in this work, we focus on the simplest scenarios in which a single molecule is coupled to either a two-mode or three-mode optical cavity. The two-mode configuration has been widely adopted in previous studies for investigating symmetry topology breaking using two sets of crossed cavities.\cite{Leonard_2017_S_p1415,Leonard_2017_N_p87,Fraxanet_2023_PRL_p263001} It also serves as a natural extension beyond the single-mode limit when considering a pair of degenerate modes with orthogonal polarizations.\cite{Moodie_2018_PRA_p33802,Fischer_2022_JCP_p34305} Few-mode cavities can also be experimentally realized using bifurcating mirrors or specialized cavity geometries such as confocal cavities.\cite{Wickenbrock_2013_PRA_p43817,Kollar_2017_NC_p14386,Vaidya_2018_PRX_p11002} For simplicity, however, we consider an experimental setup involving a planar Fabry–P\'erot cavity with an extended optical path length that supports multiple discrete modes,\cite{Simpkins_2015_AP_p1460,Chen_2024_N_p2591} among which, only two or three are assumed to be physically relevant for the reaction dynamics.  

We begin by presenting the numerical setup for simulating chemical reactions within such a cavity and then proceed to analyze results from two representative models. In the first model, the cavity-mediated reaction pathway involves a single localized vibrational transition in both the reactant and product regions. A high-order cavity mode is tuned into near-resonance with this transition, designated as the central mode with frequency $\omega_{\rm c}$. One or both adjacent modes, with frequencies $\omega_{\rm c}' = \omega_{\rm c} + \Delta$, are also included to examine the impact of neighboring modes. This scenario highlights the effects of multi-mode cavity–molecule hybridization and its implications for reaction rate modifications. In the second model, the reaction exhibits strong anharmonicity, characterized by multiple vibrational transitions with distinct energies along the reaction coordinate. In this case, by finely tuning the cavity to achieve a multi-mode strong coupling condition, where each mode is resonant with a different vibrational transition, we explore the possibility of non-additive rate enhancements mediated by multi-mode-engaged multi-photon processes.

\begin{figure}
\centering
  \begin{minipage}[c]{0.23\textwidth}
  \raggedright a1) $M=1$\au \\
    \includegraphics[width=\textwidth]{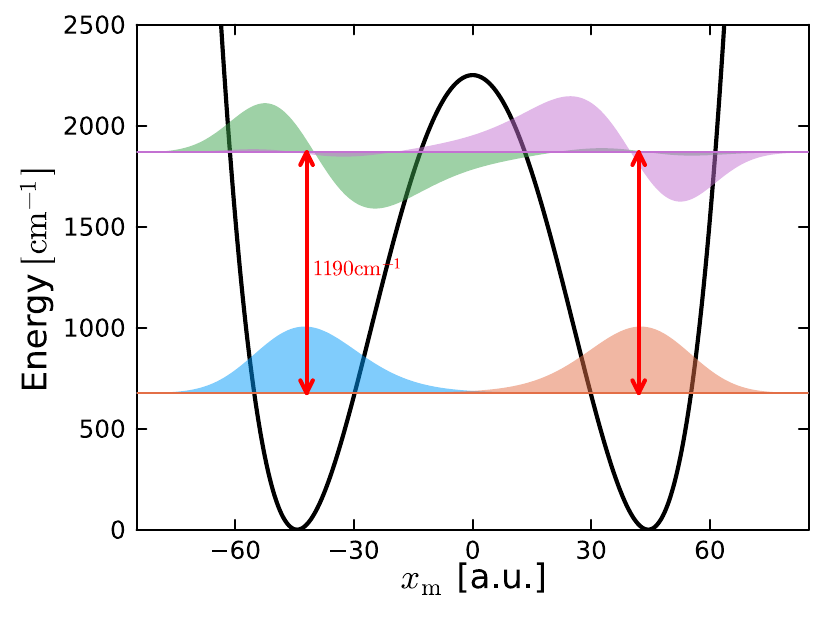}
  \end{minipage}
    \begin{minipage}[c]{0.23\textwidth}
  \raggedright b1) $M=1$\au \\
    \includegraphics[width=\textwidth]{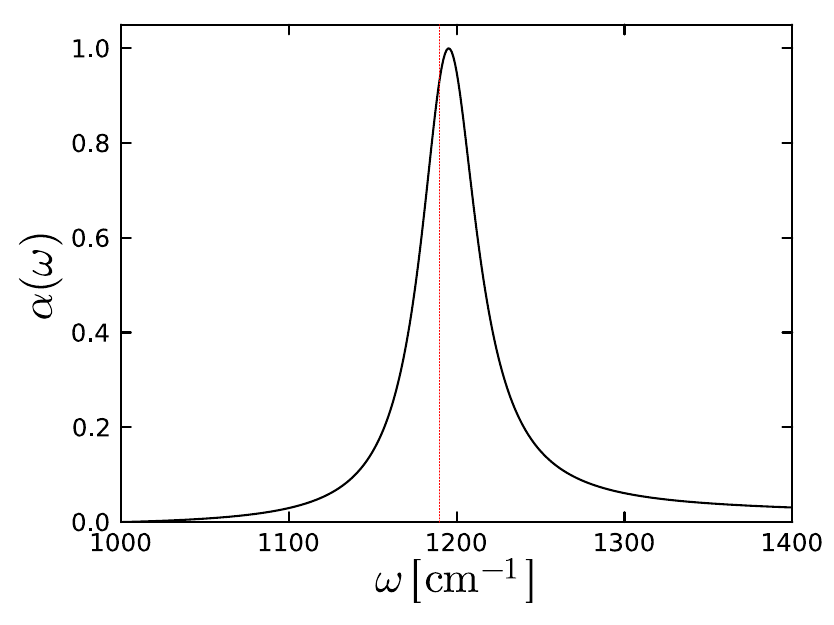}
  \end{minipage}
  \begin{minipage}[c]{0.23\textwidth}
  \raggedright a2) $M=2$\au \\
    \includegraphics[width=\textwidth]{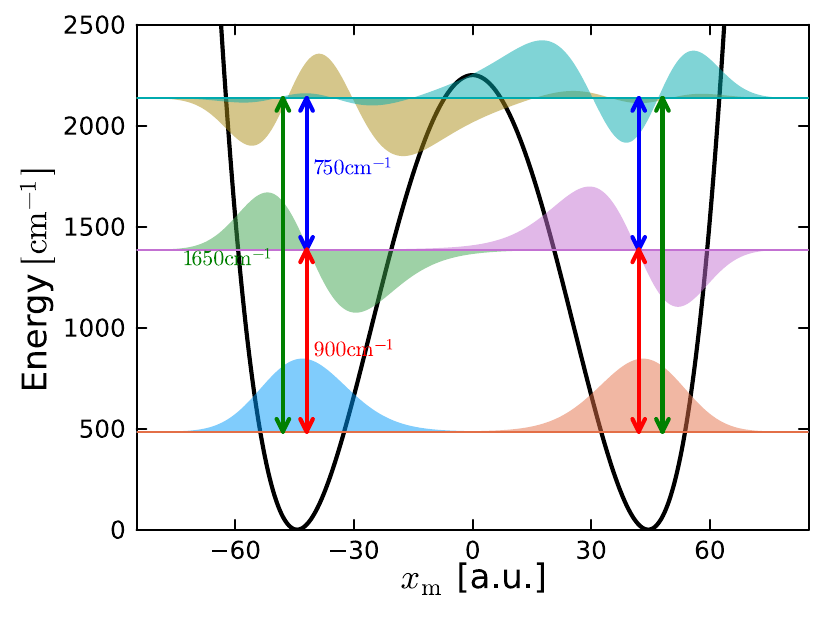}
  \end{minipage}
    \begin{minipage}[c]{0.23\textwidth}
  \raggedright b2) $M=2$\au \\
    \includegraphics[width=\textwidth]{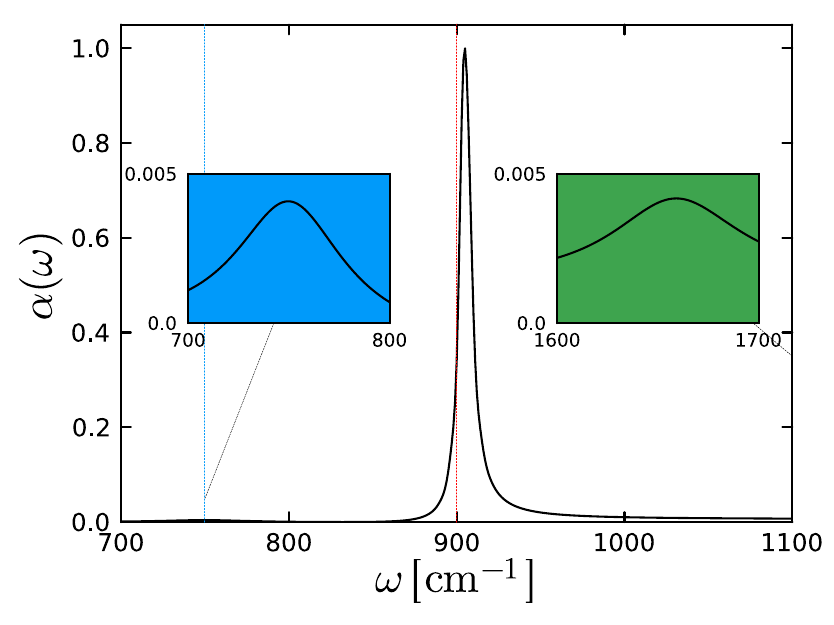}
  \end{minipage}
\caption{ a) Potential energy surface for the symmetric double-well model as defined in \Eq{PES} with $E_{\rm b}=2250\,\mathrm{cm}^{-1}$ and $a=44.4$\au for two different reduced masses $M$. The energies and the wavefunctions of localized vibrational states below the reaction barriers are shown, respectively, in each panel. b) Absorption profile of a single molecule outside the cavity for the corresponding model shown to the left. The vertical lines indicate the energy gaps of the dipole-allowed vibrational transitions in the bare molecule, as marked by the double-headed arrows in a).} \label{fig2:PES_absorption}
\end{figure}

\subsection{Numerical Details}
As an extension of previous studies to the multi-mode regime, we model the reaction using a double-well potential, consistent with earlier work:\cite{Lindoy_2023_NC_p2733,Lindoy_2024_N_p2617,Ying_2023_JCP_p84104,Hu_2023_JPCL_p11208,Ying_2024_CM_p110,Ke_J.Chem.Phys._2024_p224704,Ke_2024_JCP_p54104,Ke_J.Chem.Phys._2025_p64702}
\begin{equation}
\label{PES}
    U(x_{\rm m}) = E_{\mathrm{b}}\left(\left(\frac{x_{\rm m}}{a}\right)^2-1\right)^2,
\end{equation}
with a barrier height $E_{\rm b}$ when crossing from the minimum in the reactant well at $x_{\rm m}=-a$ to that in the product well at $x_{\rm m}=a$. This potential is illustrated in \Fig{fig2:PES_absorption}~a1) and \Fig{fig2:PES_absorption}~a2) for various parameter sets. The specific values of $E_{\rm b}$ and $a$ used in the respective model systems are provided below. 

We assume that all cavity modes share a common polarization direction $\vec{e}$. To maximize the light–matter coupling, the molecule is oriented such that its transition dipole moment aligns with the cavity field polarization vector $\vec{e}$. We further assume a linear dipole function, $\vec{\mu}(x_{\rm m}) \cdot \vec{e} = x_{\rm m}$, where the proportionality constant is absorbed into the light–matter coupling strength $\eta_{\rm i}$.

Both the molecule and the cavity modes are spanned in their respective eigenstate bases, with $|v_{\rm m}\rangle$ denoting the molecular eigenstates and $|v_i\rangle$ the eigenstates of the $i$th cavity mode. In all model systems discussed below, we retain the lowest $d_{\rm m}=12$ molecular eigenstates and the lowest $d_{\rm c}=6$ photoniv number states for each cavity mode. 

All baths are modeled using Debye–Lorentzian spectral density function:
\begin{equation}
\label{Lorentzian}
J_{\alpha}(\omega)=\frac{2\lambda_{\alpha}^2\omega\Omega_{\alpha}}{\omega^2+\Omega_{\alpha}^2},
\end{equation}
where $\Omega_{\alpha}$ is the characteristic frequency of bath $\alpha$. 
Simulations are performed at the room temperature, $T=300~{\rm K}$. Unless otherwise specified, the molecular bath parameters are fixed at $\lambda_{\rm m}=100~{\rm cm}^{-1}$ and $\Omega_{\rm m}=200~{\rm cm}^{-1}$, while all cavity baths are assigned $\Omega_{\rm c}=1000~{\rm cm}^{-1}$ and $\lambda_{\rm c}=100~{\rm cm}^{-1}$. The bath correlation function in \Eq{timecorrelation} is expanded via the Pad\'e decomposition scheme with $P=4$ terms. Each effective bath mode is represented using $d_{\rm e}=10$ states. Convergence is ensured by checking the results with respect to the time step $\delta t$ and the maximal bond dimension $D_{\rm max}$ in the TTNS decomposition.  

The rate constant for the forward reaction from the reactant region to the product region, separated by a dividing surface at $x_{\rm m}^{\ddagger}=0$, is computed using the flux-side correlation formalism \cite{Miller_1983_JCP_p4889,Craig_2007_JCP_p144503,Chen_2009_JCP_p134505,Ke_2022_JCP_p34103}
\begin{equation}
k=\lim_{t\rightarrow t_{\rm plateau}} \frac{C_F(t)}{1-(1+1/K)P_{\rm p}(t)}.
\end{equation}
The flux-side correlation function is given by $C_F(t)=\langle {\rm II}|\tilde{F}|\Psi(t)\rangle$, which corresponds to the expectation value of the flux operator $F$ when the molecule is initially prepared in the reactant region.  The flux operator is defined as $F=i[H, h]$ and $h=\theta(x_{\rm m}-x_{\rm m}^{\ddagger})$ is the Heaviside step function projecting onto the product region (i.e., $h=1$ for $x\ge x_{\rm m}$ and $h=0$ otherwise). Here, the tracing operation is carried out by contracting with the reference state $|{\rm II}\rangle=|1_{\mathrm{sys}}\rangle\otimes |\mathbf{n=0}\rangle$, where $|1_{\mathrm{sys}}\rangle=\otimes_s\sum_{v^{}_s=v'_s}|v^{}_sv'_s\rangle$, is the unit vector in twin space, and the environmental modes reside in the vacuum Fock state. 
The time-dependent population in the product region is given by $P_{\rm p}(t) = \langle {\rm II}|\hat{h}|\Psi(t)\rangle$, and its equilibrium value, $P_{\rm p}^{\rm eq}$, is obtained through the imaginary-time HEOM method.\cite{Tanimura_J.Chem.Phys._2014_p44114,ke2025non} The equilibrium constant is then $K=P_{\rm p}^{\rm eq}/(1-P_{\rm p}^{\rm eq})$, presenting the ratio of product to reactant populations at equilibrium.  The plateau time $t_{\rm plateau}$ marks the onset of the kinetic regime, where transient dynamics have subsided and the condensed-phase reactive dynamics are governed by a well-defined rate process, such that $k(t)$ reaches a stationary value. The molecular absorption profile is obtained by performing the Fourier transform of the dipole-dipole autocorrelation function, $\alpha(\omega)\propto \frac{1}{2\pi}\int_{-\infty}^{\infty}\mathrm{d}t e^{i\omega t}\langle \vec{\mu}(x,t)\vec{\mu}(x,0)\rangle$.

Throughout this study, we denote the reaction rates in single-mode, two-mode, or three-mode cavities as $k_{\rm c}^1$, $k_{\rm c}^2$, and $k_{\rm c}^3$, respectively, and the rate outside the cavity as $k_{\rm o}$.
  
\begin{figure*}
\centering
 \begin{minipage}[c]{0.24\textwidth}
    \raggedright a) single-mode cavity
    \includegraphics[width=\textwidth]{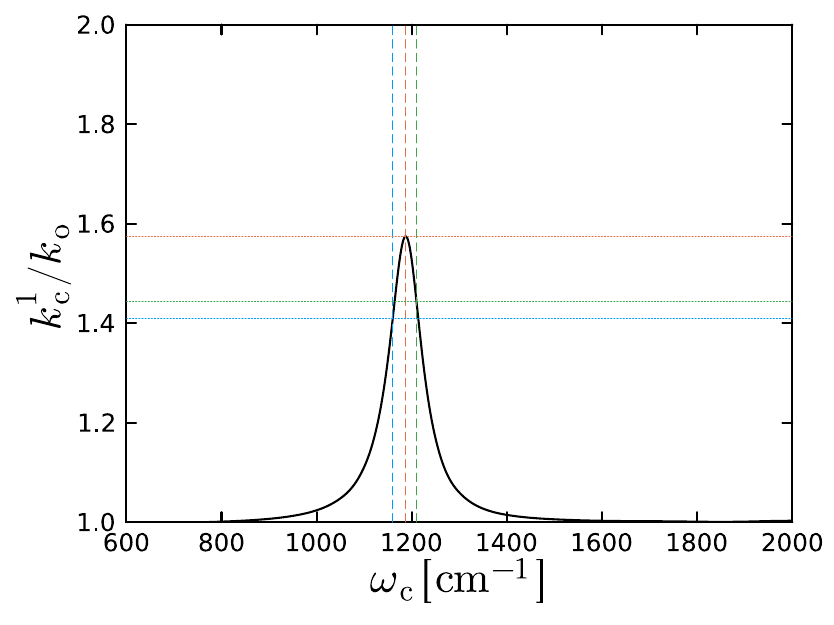}
 \end{minipage}
 \begin{minipage}[c]{0.24\textwidth}
    \raggedright  b) $\omega_c=1160\,\mathrm{cm}^{-1}$
    \includegraphics[width=\textwidth]{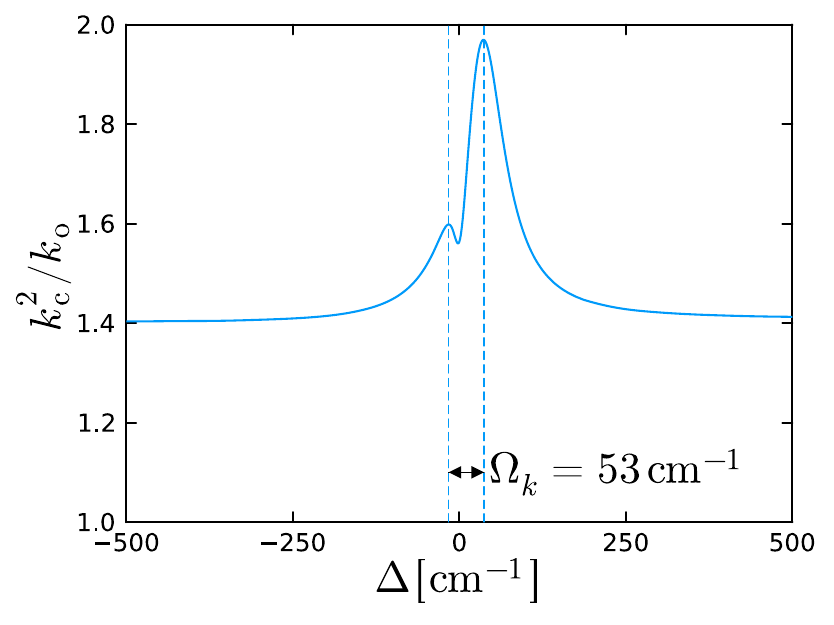}
  \end{minipage}
   \begin{minipage}[c]{0.24\textwidth}
    \raggedright  c) $\omega_c=1185\,\mathrm{cm}^{-1}$
    \includegraphics[width=\textwidth]{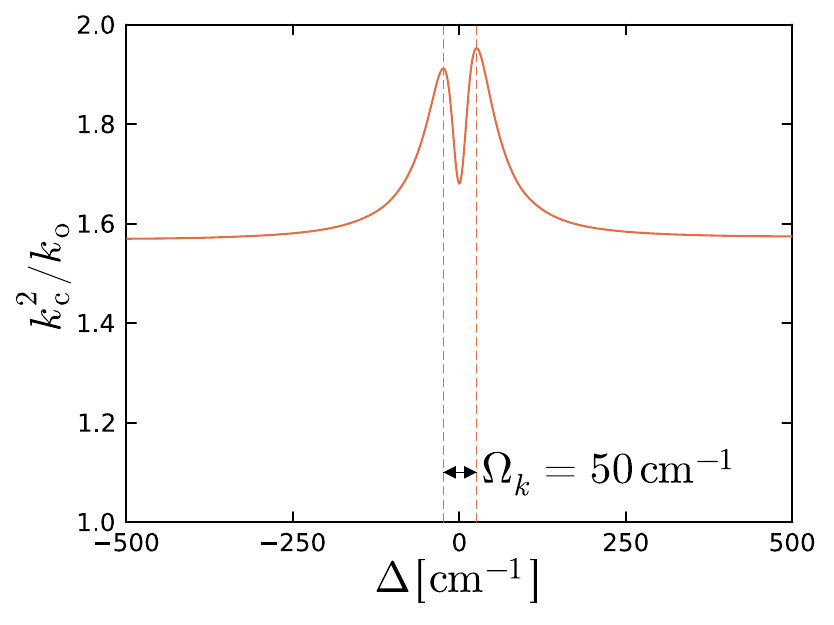}
  \end{minipage}
   \begin{minipage}[c]{0.24\textwidth}
    \raggedright  d) $\omega_c=1210\,\mathrm{cm}^{-1}$
    \includegraphics[width=\textwidth]{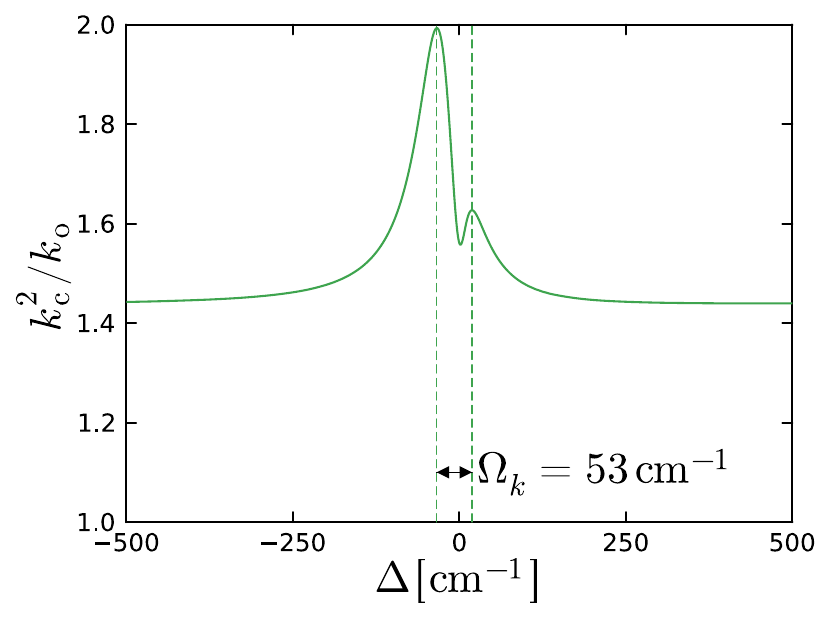}
  \end{minipage}
\caption{a) Rate modification (ratio $k^1_{\mathrm c}/k_{\mathrm{o}}$) for Model I in a single-mode cavity as a function of the cavity frequency $\omega_{\mathrm{c}}$. b-c) Rate modification ((ratio $k^2_{\mathrm c}/k_{\mathrm{o}}$)) for Model I as a function of the FSR $\Delta$ in a two-mode cavity, where $\omega_{\mathrm c}$ is the frequency for the central cavity mode, and $\omega'_{\mathrm c}=\omega_{\mathrm c}+\Delta$ for a neighboring mode. The light-matter coupling strength is set to $\eta_1=\eta_2=0.00125~$\au} \label{fig3:rates_model1}
\end{figure*}
\begin{figure}
\centering
 \begin{minipage}[c]{0.23\textwidth}
    \raggedright  a1) \\$\omega_c=1160\,\mathrm{cm}^{-1}$
    \includegraphics[width=\textwidth]{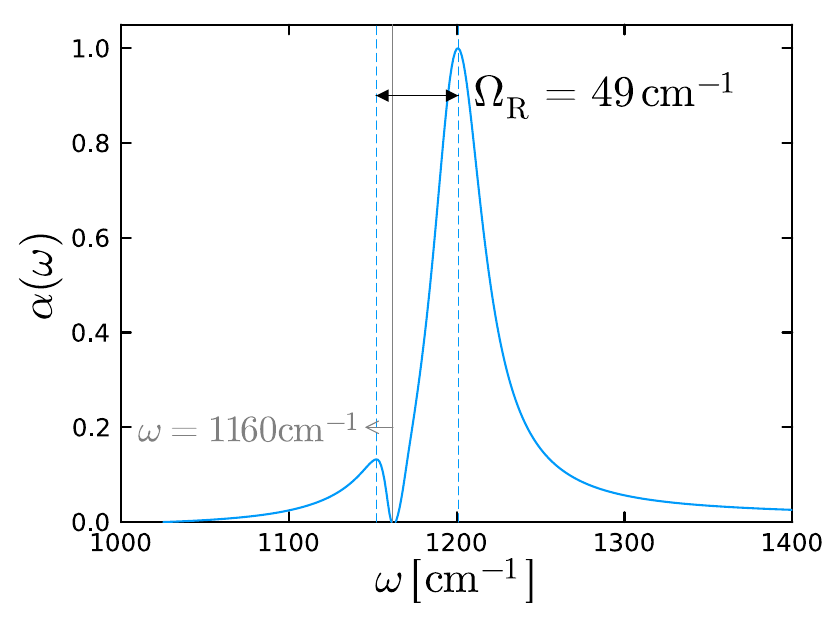}
 \end{minipage}
 \begin{minipage}[c]{0.23\textwidth}
    \raggedright a2) $\omega_c=1185\,\mathrm{cm}^{-1},\omega'_c=1160\,\mathrm{cm}^{-1}$
    \includegraphics[width=\textwidth]{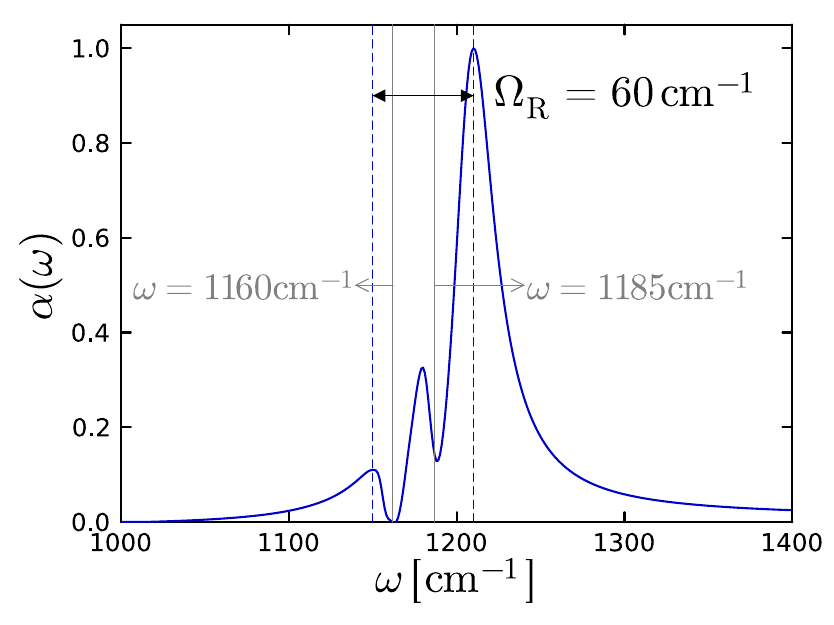}
 \end{minipage}
 \begin{minipage}[c]{0.23\textwidth}
    \raggedright  b1) \\$\omega_c=1185\,\mathrm{cm}^{-1}$
    \includegraphics[width=\textwidth]{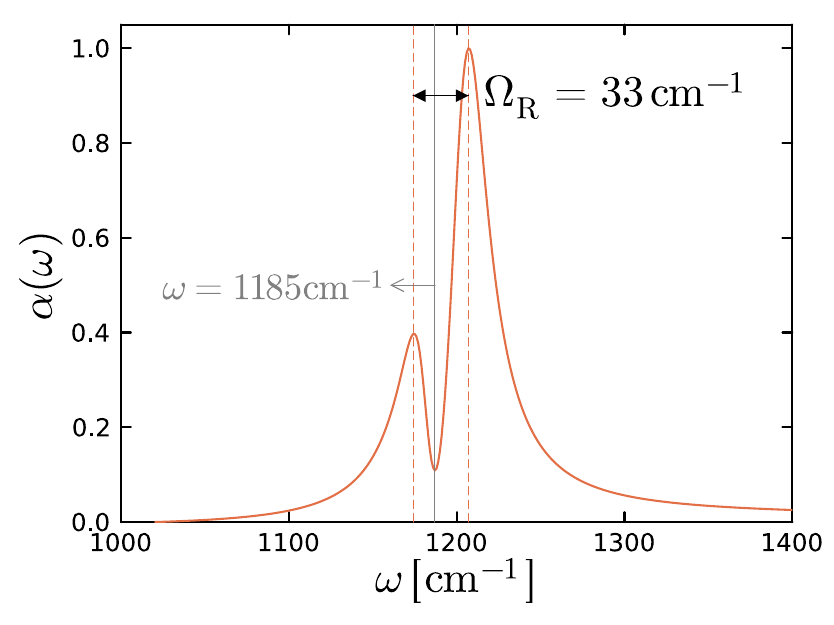}
 \end{minipage}
 \begin{minipage}[c]{0.23\textwidth}
    \raggedright  b2) $\omega_c=1185\,\mathrm{cm}^{-1},\omega'_c=1185\,\mathrm{cm}^{-1}$
    \includegraphics[width=\textwidth]{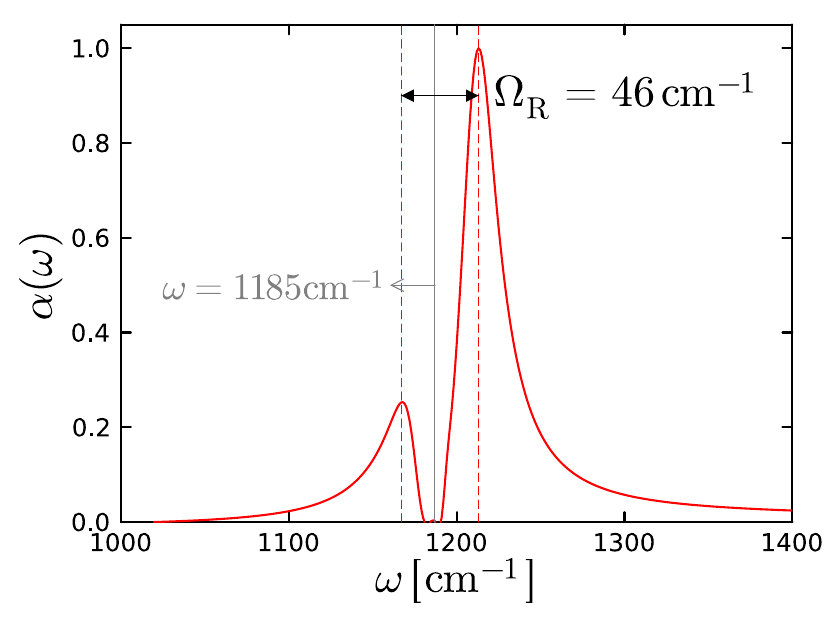}
 \end{minipage}
 \begin{minipage}[c]{0.23\textwidth}
    \raggedright  c1) \\$\omega_c=1210\,\mathrm{cm}^{-1}$
    \includegraphics[width=\textwidth]{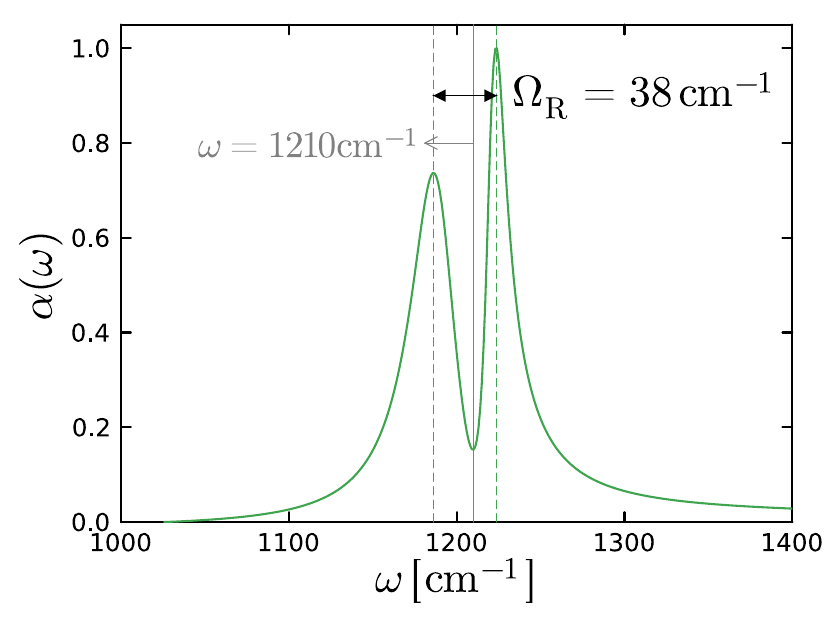}
 \end{minipage}
  \begin{minipage}[c]{0.23\textwidth}
    \raggedright c2) $\omega_c=1185\,\mathrm{cm}^{-1},\omega'_c=1210\,\mathrm{cm}^{-1}$
    \includegraphics[width=\textwidth]{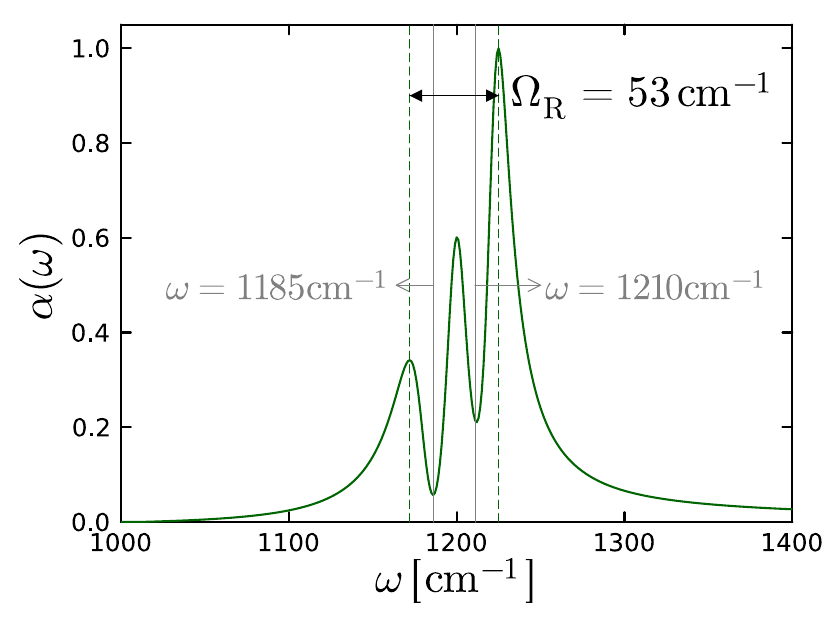}
 \end{minipage}
\caption{Absorption profiles of a single molecule (Model I) in a single-mode cavity (left column) with three different cavity frequencies, and a two-mode cavity (right column) with a central cavity frquency at $\omega_{\mathrm{c}}=1185\mathrm{cm}^{-1}$ and three different neighboring cavity frequencies $\omega'_{\rm c}$. The light-matter coupling strength is set to $\eta_1=\eta_2=0.00125~$\au} \label{fig4:absorption}
\end{figure}

\begin{figure*}
\centering
  \begin{minipage}[c]{0.35\textwidth}
  \raggedright a)
    \includegraphics[width=\textwidth]{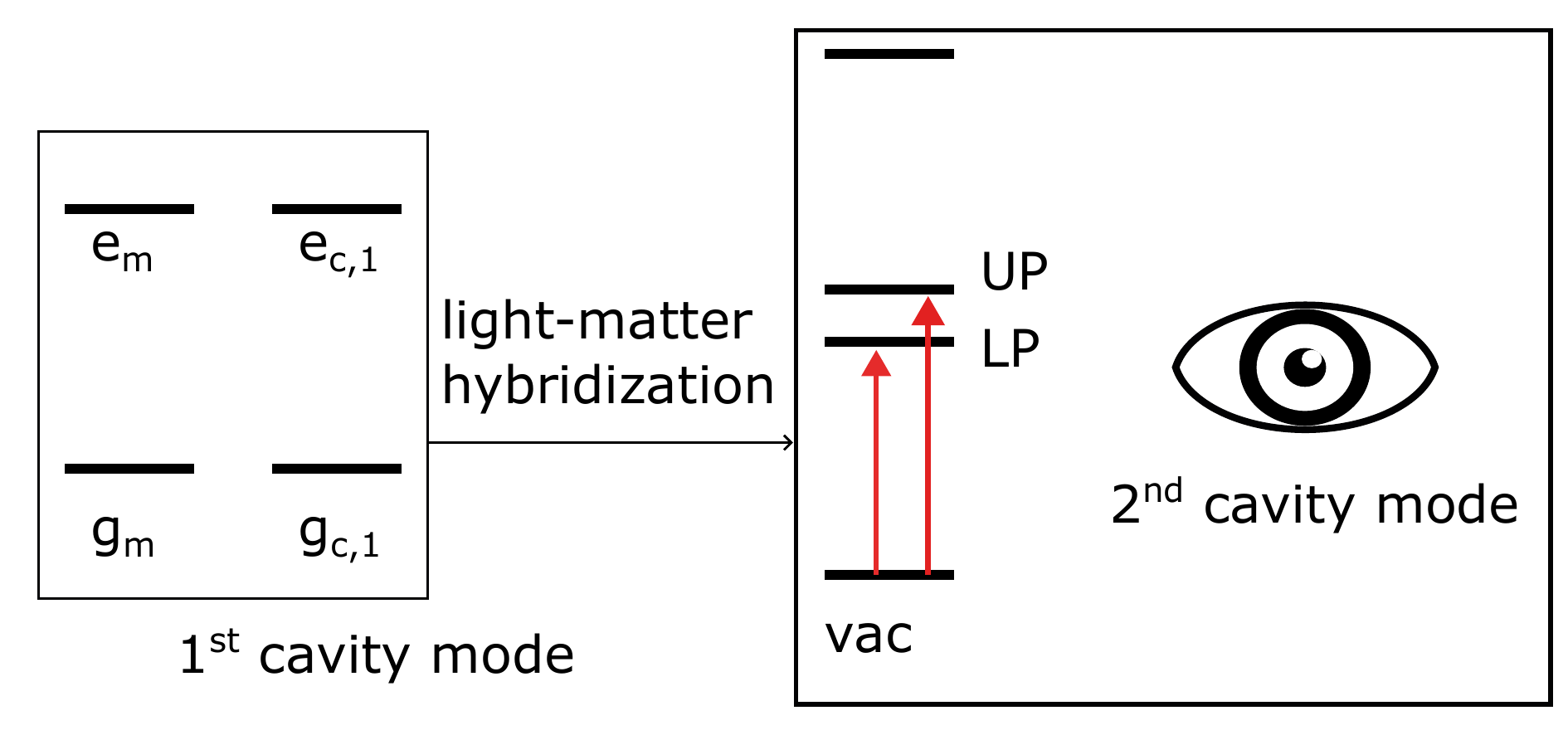}
  \end{minipage} 
    \begin{minipage}[c]{0.6\textwidth}
    \raggedright b)
    \includegraphics[width=\textwidth]{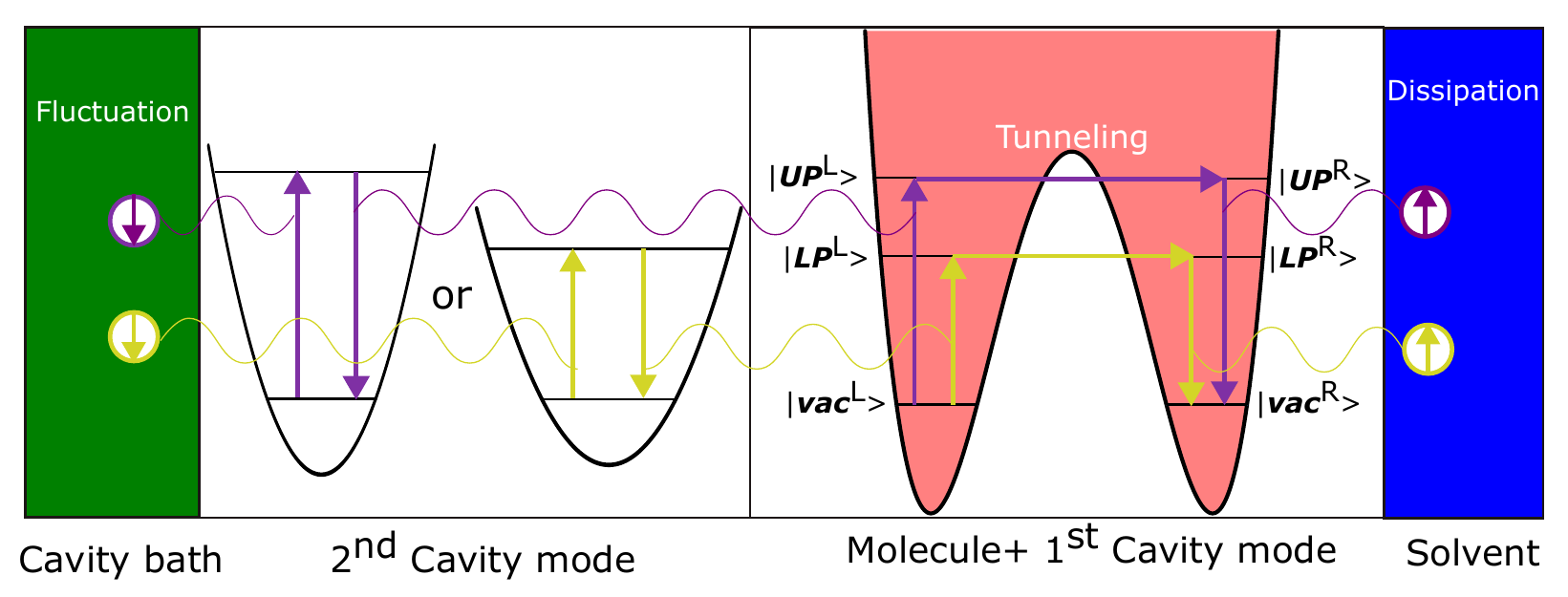}
  \end{minipage} 
\caption{a) Schematic illustration of the hybridization between a molecular transition and the first cavity mode, forming two polaritonic states, which modifies the resonant conditions for the second cavity mode. b) Reaction mechanism in a two-mode cavity, which is responsible for the additional rate enhancement observed in \Fig{fig3:rates_model1}~b)-d).} \label{fig5:schematic}
\end{figure*}

\begin{figure}
\centering
\begin{minipage}[c]{0.4\textwidth}
\raggedright a)
    \includegraphics[width=\textwidth]{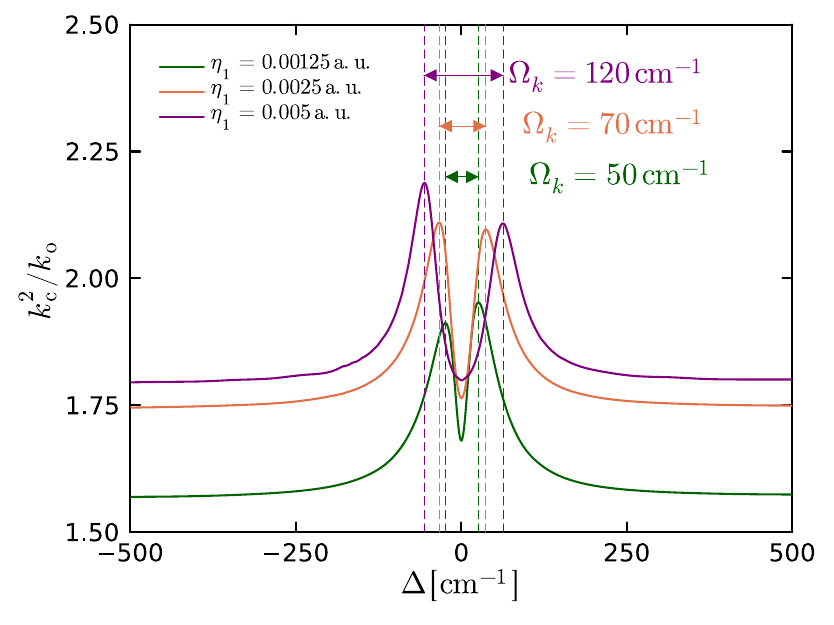}
  \end{minipage}
  \begin{minipage}[c]{0.4\textwidth}
  \raggedright b)
    \includegraphics[width=\textwidth]{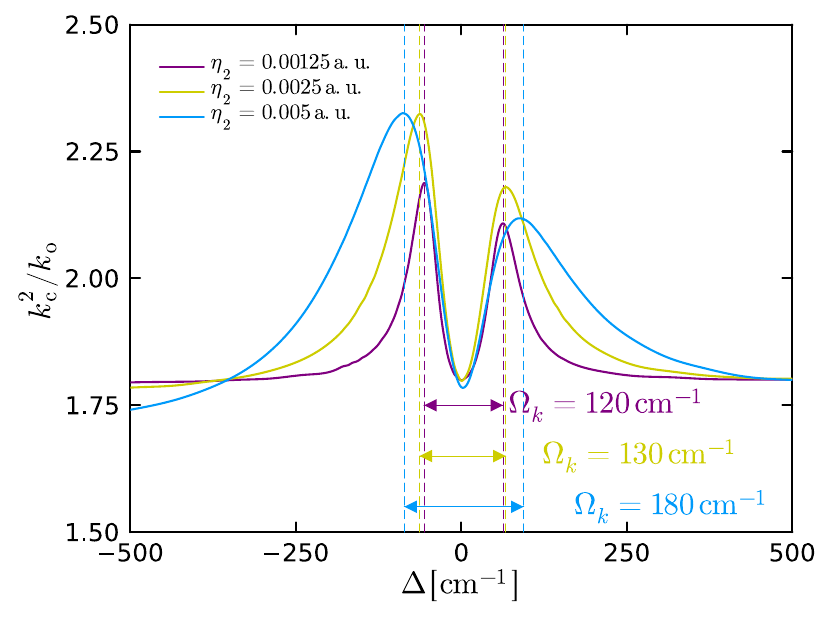}
  \end{minipage}
\caption{Rate modification factor in a two-mode cavity, $(k_{\mathrm c}^2/k_{\mathrm{o}})$, for Model I as a function of the FSR $\Delta$ for different light-matter coupling strengths.  In panel a), the results are for a fixed $\eta_2=0.00125$\au and varying $\eta_1$. In panel b), we keep $\eta_1=0.005$\au and vary $\eta_2$. The central cavity frequency is $\omega_{\mathrm{c}}=1185\,\mathrm{cm}^{-1}$.} \label{fig6:etac}
\end{figure}

\begin{figure}
\centering
  \begin{minipage}[c]{0.4\textwidth}
    \includegraphics[width=\textwidth]{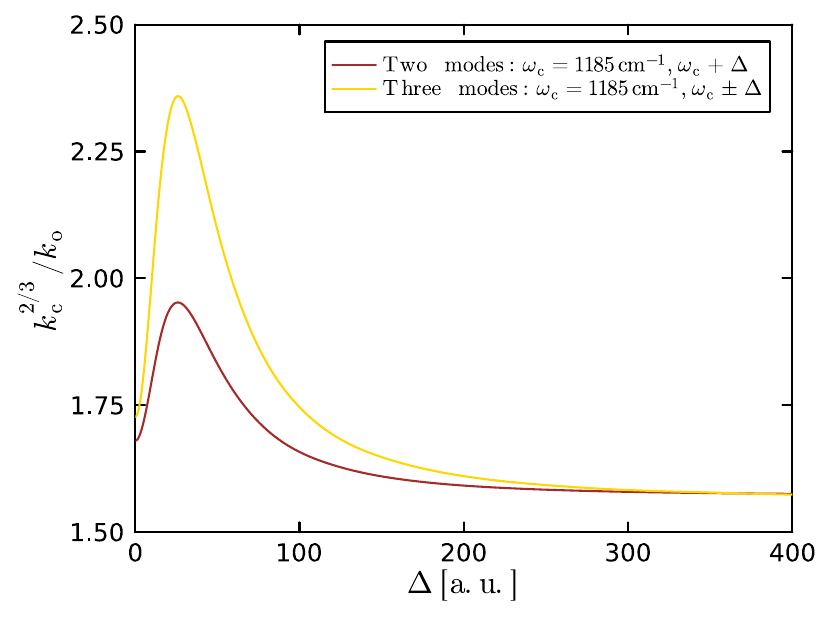}
  \end{minipage}
\caption{Comparison of the rate modification factor of Model I in a two-mode cavity ($k_{\mathrm c}^2/k_{\mathrm{o}}$) and a three-mode cavity ($k_{\mathrm c}^3/k_{\mathrm{o}}$). The results are displayed as a function of the positive FSR $\Delta$. In both scenarios, we fix a central cavity mode at $\omega_{\mathrm{c}}=1185\,\mathrm{cm}^{-1}$. In the two-mode case, the frequency of the second mode varies with $\Delta$ as $\omega'_{\mathrm{c}}=\omega_{\mathrm{c}}+\Delta$.  In the three-mode scenario, the nearest-neighboring modes take the frequency of $\omega_{\mathrm{c}}\pm \Delta$, respectively. The light-matter coupling strength is fixed at $\eta_1=\eta_2=\eta_3=0.00125~$\au} \label{fig7:3modes}
\end{figure}

\subsection{\label{subsec:modelI}Model I}
We begin our analysis with a symmetric double-well potential energy surface characterized by a barrier height of $E_{b}=2250\,\mathrm{cm}^{-1}$ and a width parameter $a=44.4$\au The reduced mass of the reactive bond is set to $M=1$\au This model has been employed in previous studies to investigate reaction dynamics in a single-mode optical cavity.\cite{Lindoy_2023_NC_p2733,Lindoy_2024_N_p2617,Ying_2023_JCP_p84104,Hu_2023_JPCL_p11208,Ying_2024_CM_p110,Ke_J.Chem.Phys._2024_p224704,Ke_2024_JCP_p54104,Ke_J.Chem.Phys._2025_p64702} 

Notably, two distinct dipole-allowed vibrational transitions contribute to the cavity-induced resonant rate modification. The first involves a transition between the eigenstates $|v_m=0\rangle$ and $|v_m=3\rangle$, with a transitional energy of $\delta E_{0\leftrightarrow 3}=1238\,\mathrm{cm}^{-1}$. The second transition occurs between adjacent states $|v_m=1\rangle$ and $|v_m=2\rangle$ with $\delta E_{1\leftrightarrow 2}=1140\,\mathrm{cm}^{-1}$. When the broadening caused by the coupling to the solvent bath (denoted by $\lambda_{\rm m}$) is sufficiently weak, these transitions appear as two well-resolved peaks in the rate enhancement profiles ($k_{\rm c}^1/k_{\rm o}$) as a function of the cavity frequency $\omega_{\rm c}$ in a single-mode cavity. As $\lambda_m$ increases, these peaks broaden and eventually coalesce into a single resonant feature, as demonstrated in the SI. 
The position of the merged peak lies between the two vibrational transition energies and depends on the cavity loss strength $\lambda_{\rm c}$ and the light-matter coupling strength $\eta_{\rm c}$. In this regime, the molecular system can be effectively described by the localized states formed via symmetric and antisymmetric superpositions of the bare eigenstates:
\begin{equation}
\begin{split}
   |0^L\rangle=\frac{|v_{\rm m}=0\rangle +|v_{\rm m}=1\rangle}{\sqrt{2}},  \quad |0^R\rangle=\frac{|v_{\rm m}=0\rangle -|v_{\rm m}=1\rangle}{\sqrt{2}}, \\
   |1^L\rangle=\frac{|v_{\rm m}=2\rangle +|v_{\rm m}=3\rangle}{\sqrt{2}}, \quad |1^R\rangle=\frac{|v_{\rm m}=2\rangle -|v_{\rm m}=3\rangle}{\sqrt{2}},
   \end{split}
\end{equation}
as illustrated in \Fig{fig2:PES_absorption}~a1).
For $\lambda_{\rm m}=100\,\mathrm{cm}^{-1}$, the single peak observed in the molecular absorption profile outside the cavity (see \Fig{fig2:PES_absorption}~b1)), as well as the resonance feature in the rate enhancement profile $k_{\rm c}^1/k_{\rm o}$ at $\eta_{\rm c}=0.00125$\au (see \Fig{fig3:rates_model1}~a)), can both be attributed to the degenerate transitions $|0^{L}\rangle \leftrightarrow |1^{L}\rangle$ in the reactant well and $|0^{R}\rangle \leftrightarrow |1^{R}\rangle$ in the product well.

We now extend the analysis to a two-mode cavity. We consider a setup in which the central cavity mode with frequency $\omega_{\rm c}$ is tuned near resonance with a specific molecular vibrational transition. A neighboring mode, with frequency $\omega_{\rm c}'=\omega_{\rm c}+\Delta$, is offset by the FSR $\Delta$,  which varies for cavities with different thickness $L$. In particular, the case of $\Delta=0$ does not necessarily imply an infinitely extended cavity with widely separated mirrors, but rather corresponds to a pair of degenerate optical modes that differ only in polarization--supported by the same cavity geometry. 

\Fig{fig3:rates_model1}~b)-d) present how the reaction rate enhancement $k_{\rm c}^2/k_{\rm o}$ evolves in a two-mode cavity as a function of the tunable free spectral range $\Delta$,\cite{Kollar_2015_NJP_p43012} for three different central cavity frequencies near the resonant vibrational transition.
The corresponding frequency positions are indicated by vertical lines in \Fig{fig3:rates_model1}~a). The light–matter coupling strengths are held constant at $\eta_{\rm 1}=\eta_{2}=0.00125$\au, consistent with previous studies suggesting that the light-matter coupling strength is largely independent of the cavity length in Fabry-P\'erot geometries.\cite{Simpkins_2015_AP_p1460, Hertzog_2020_C_p612} Although the energy dissipation has been reported to decrease with increasing cavity length $L$ (i.e., decreasing $\Delta$),\cite{Hertzog_2020_C_p612} we keep $\lambda_{\rm m}$ and  $\lambda_{\rm c}$ fixed to ensure fair comparison.  These parameters play a crucial role in shaping the reaction dynamics, and varying them could significantly complicate the interpretation.\cite{Ke_J.Chem.Phys._2025_p64702}

When the FSR is sufficiently large ($\Delta>200\,\mathrm{cm}^{-1}$), as in high-finesse cavities, the cavity modes are well-separated in frequency, and the reaction rates are essentially indistinguishable from the single-mode case. This behavior confirms that far-off-resonant cavity modes exert negligible influence on the reaction dynamics. 

However, when the cavity optical path length $L$ is extended such that the FSR $\Delta$ becomes comparable to the Rabi splitting, qualitatively new phenomena emerge due to coherent interactions between closely spaced cavity modes and the molecular transition.  Specifically, the inclusion of a nearby cavity mode in energy leads to enhanced rates that surpass the single-mode case. In the regime $|\Delta|<200\,\mathrm{cm}^{-1}$, the enhancement factor $k_{\rm c}^2/k_{\rm o}$ exhibits a characteristic splitting as a function of $\Delta$, with a peak-to-peak separation of approximately $\Omega_k\approx 50\,\mathrm{cm}^{-1}$ and a local minimum at $\Delta=0$. The optimal FSR that yields the maximum enhancement depends sensitively on the central frequency $\omega_{\rm c}$. For example, when $\omega_{\rm c}=1160\,\mathrm{cm}^{-1}$, slightly below the vibrational transitional energy $\delta E$ (or more accurately the peak position in \Fig{fig3:rates_model1}~a)), a higher-frequency neighboring mode ($\Delta>0)$ contributes most significantly, producing a dominant peak at $\Delta=35\mathrm{cm}^{-1}$, as shown in \Fig{fig3:rates_model1} b). In contrast, when $\omega_{\rm c}=1210\,\mathrm{cm}^{-1}$, above $\delta E$, a stronger rate enhancement is observed for a lower-frequency neighboring mode ($\Delta=-35\,\mathrm{cm}^{-1}$), along with a weaker secondary peak at $\Delta=18\,\mathrm{cm}^{-1}$, as seen in \Fig{fig3:rates_model1} d).  For $\omega_{\rm c}=1185\,\mathrm{cm}^{-1}$, which corresponds to the peak position in \Fig{fig3:rates_model1}~a), the profile of $k_{\rm c}^2/k_{\rm o}$ as a function of $\Delta$ exhibits an almost symmetric splitting, with the neighboring modes of higher and lower frequencies contributing comparably to the rate enhancement, as displayed in \Fig{fig3:rates_model1}~c).

This splitting in the rate modification profile (i.e., $k_{\rm c}^2/k_{\rm o}$ versus $\Delta$) is reminiscent of the polariton doublet observed in the absorption spectrum of a molecule coupled to a single-mode cavity [see \Fig{fig4:absorption}~a1), b1), and c1)]. In the single-mode scenario, the resonant light–matter coupling leads to the formation of upper and lower polaritons, which manifest as two distinct peaks in the absorption spectrum. In a two-mode cavity, the additional mode plays a role analogous to that of a probing beam in linear absorption measurements. Specifically, the central cavity mode couples resonantly with a molecular vibrational transition to generate two polaritonic states. These hybridized light-matter states, both of which retain significant characteristics of the vibrationally excited state, subsequently alter the resonance conditions perceived by the second cavity mode, as schematically illustrated in \Fig{fig5:schematic}~a). In the context of the double-well model as depicted in \Fig{fig2:PES_absorption}~a1), the coupling of the central cavity mode with the transitions $|0^L\rangle \rightarrow |1^L\rangle$ and $|0^R\rangle \rightarrow |1^R\rangle$ gives rise to two sets of upper and lower polaritonic states: $|{\rm LP}^{L/R}\rangle$ and $|{\rm UP}^{L/R}\rangle$.  When a neighboring cavity mode is present, it can resonantly interact with these polaritonic transitions, effectively opening up two distinct cavity-induced intramolecular reaction pathways, as schematically illustrated in \Fig{fig5:schematic}~b). As a result, the overall reaction rate can exceed that observed in the single-mode limit. The rate enhancement is maximized when the frequency of the second mode, $\omega_{\rm c}'$, aligns with the transition energy from the vacuum (hybridized ground) state to one of the polaritonic states, which retain an appreciable molecular character.

This explanation is further supported by analyzing how the rate modification responds to the variations in the coupling strengths $\eta_1$ and $\eta_2$. \Fig{fig6:etac}~a) displays the rate modification ratio $k_{\rm c}^2/k_{\rm o}$ as a function of the FSR  $\Delta$ for three different coupling strengths to the central cavity mode $\eta_1$, while keeping the coupling to the neighboring cavity mode $\eta_{2}=0.00125$\au as a constant. The central cavity frequency is set at $\omega_{\rm c}=1185\,\mathrm{cm}^{-1}$. As $\eta_1$ increases, the energy gap between the lower and upper polaritons--formed by the admixing of the vibrational excited states $|1^{L/R}\rangle$ and the central cavity mode--widens. This shifts the optimal FSR (or equivalently, the neighboring frequency) that maximizes the reaction rate. Consequently, the peak-to-peak splitting $\Omega_k$ in $k_{\rm c}^2/k_{\rm o}$ increases with $\eta_1$, while the peak widths remain relatively unaffected.

In \Fig{fig6:etac}~b), we fix $\eta_1=0.005$\au and vary the coupling strength $\eta_2$ to compute $k_{\rm c}^2/k_{\rm o}$ as a function of the FSR $\Delta$. 
In all cases, the resulting splitting $\Omega_k$ exceeds the energy difference between the two dipole-allowed eigenstate transitions ($|v_{\rm m}=0\rangle \leftrightarrow |v_{\rm m}=3\rangle$ and $|v_{\rm m}=1\rangle \leftrightarrow |v_{\rm m}=2\rangle$). This observation rules out the possibility that the appearance of two peaks in ($k_{\rm c}^2/k_{\rm o}$ versus $\Delta$) stems from a reverse process of broadening-induced peak merging seen in the single-mode profile ($k_{c}^1/k_{\rm o}$ versus $\omega_{\rm c}$).  As shown in \Fig{fig6:etac}~b), increasing the coupling to the neighboring mode $\eta_2$ not only shifts the peak positions further apart--contributing to a larger splitting--but also leads to notable broadening of the two resolved peaks. This effect explains the quantitative discrepancy between the splitting $\Omega_{k}$ observed in the two-mode rate modification profile [\Fig{fig3:rates_model1}~c)] and the single-mode Rabi splitting evident in the molecular absorption spectrum [\Fig{fig2:PES_absorption}~b1)].  

However, it is important to emphasize that both cavity modes interact simultaneously with the molecule. \Fig{fig4:absorption}~a2), b2), and c2) show the molecular absorption profiles in a two-mode cavity. In the degenerate case—where the two cavity modes are energetically identical—the effective Rabi splitting between the bright polaritonic branches exceeds that of the single-mode scenario. For example, as illustrated in \Fig{fig4:absorption}~b1) and b2), the Rabi splitting increases from $\Omega_{\rm R}=33\,\mathrm{cm}^{-1}$ in the single-mode case to $\Omega_{\rm R}=46\,\mathrm{cm}^{-1}$ in the degenerate two-mode configuration. More intriguingly, in non-degenerate two-mode systems, a single vibrational transition can hybridize with both cavity modes, resulting in the formation of three polaritonic states. This gives rise to a distinctive three-peak structure in the absorption spectra, as seen in \Fig{fig4:absorption}~a2) and c2). The emergence of a mid-polariton state is consistent with recent theoretical predictions presented in Ref.~\onlinecite{Godsi_2023_JCP_p134307}.

Increasing the cavity damping strength $\lambda_{\rm c}$ induces spectral broadening of the polaritonic resonances. Strong dissipation blurs the distinction between the hybridized light–matter states: the polaritonic peaks broaden, overlap, and merge into an effectively degenerate manifold.  In this regime, the subtle fine structure arising from hybridization with one cavity mode is no longer resolvable by the other, erasing the peak splitting that signals multi-mode strong coupling. Further details are given in the SI. 

Up to this point, we have considered only one neighboring cavity mode. However, in realistic Fabry–P\'erot cavities, the discrete cavity modes are uniformly spaced. For a specified central cavity mode with frequency $\omega_{\rm c}$, the nearest modes appear symmetrically at $\omega'_{\rm c} = \omega_{\rm c} \pm \Delta$,\cite{Herrera_2024_PTA_p20230343} where we assume $\Delta$ is positive here. In \Fig{fig7:3modes}, we compare the rate modification as a function of the FSR $\Delta$ for two-mode and three-mode cavity configurations, both with the central mode fixed at $\omega_{\rm c}=1185\,\mathrm{cm}^{-1}$. In the two-mode case, only the higher-frequency neighboring mode is included, while in the three-mode setup, both adjacent modes detuned by $\pm \Delta$ are included. The light-matter coupling strengths are held constant at $\eta_{\rm 1}=\eta_2=\eta_3=0.00125$\au Both configurations exhibit a turnover in $k_{\rm c}^i/k_{\rm o}$ as $\Delta$ increases, with a maximum occurring at $\Delta=35\,\mathrm{cm}^{-1}$--approximatley equal to the single-mode Rabi splitting. Notably, the three-mode cavity achieves a higher peak rate enhancement, indicating cooperative effects when more cavity modes participate. This result highlights the potential of utilizing multi-mode cavities in further optimizing reaction kinetics.  

In short, our first model system underscores the capability of multi-mode cavities to boost chemical reactivities beyond what is achievable in single-mode scenarios. This effect is especially pronounced in long-path cavities with small FSRs, which have been used in practice for novel comb generation\cite{Erickson_2014_PRL_p187002} and polaritonic chemistry experiments.\cite{Simpkins_2015_AP_p1460,Chen_2024_N_p2591} These cavities, which are reported to reduce dissipation, may also enable multiple cavity modes to lie near resonance with a molecular vibrational transition. By tuning the cavity length $L$ such that a high-order cavity mode is brought into resonance with the vibrational transition and the FSR is comparable to the Rabi splitting, neighboring modes can synergistically amplify the reaction rate. In passing, the observed splitting in the rate modification profiles as a function of $\Delta$ in two-mode cavities cannot be captured by simple analytical rate theories such as Fermi’s Golden Rule. This is discussed in more detail in the SI, emphasizing the necessity of an explicit quantum mechanical treatment of hybrid light-matter states in the study of vibrational polariton chemistry.

Note in passing that in most of the cases and parameter regimes we have examined, the inclusion of a second cavity mode leads to an overall enhancement of the reaction rate.  However, while the prevailing trend is rate enhancement upon adding a second cavity mode, there do exist regimes--particularly under strong light–matter coupling in the weak cavity damping limit--where the opposite occurs, i.e., the reaction rate is reduced by the presence of the additional cavity mode. Examples are provided in the SI. 

While the prevailing effect of including additional cavity modes is to enhance the reaction rate across the majority of parameter regimes considered, we also identify exceptions. In particular, under strong light–matter coupling and weak cavity damping, the presence of a second mode instead suppresses the rate enhancement. Examples of this behavior are provided in the SI.

\subsection{\label{subsec:modelI}Model II}
\begin{figure*}
\centering
 \begin{minipage}[c]{0.24\textwidth}
    \raggedright a) single-mode cavity
    \includegraphics[width=\textwidth]{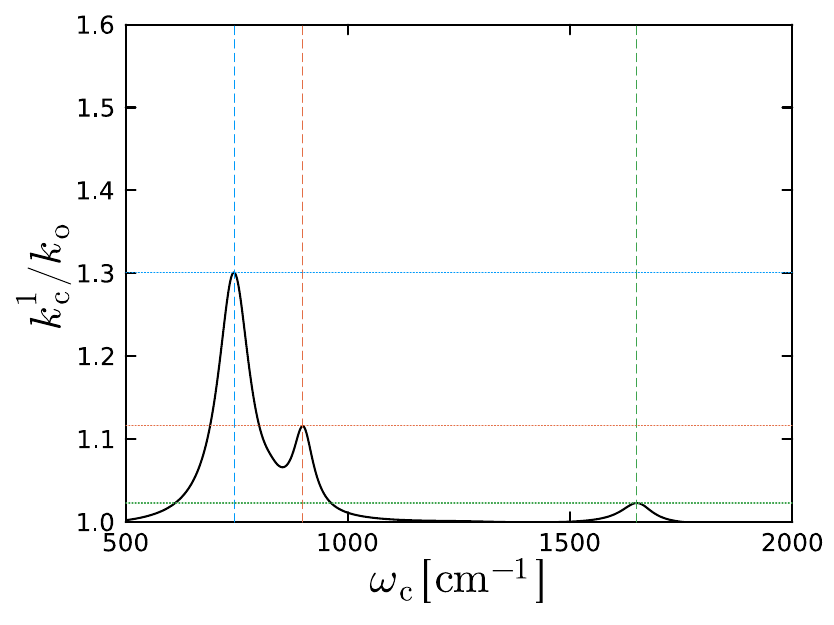}
 \end{minipage}
 \begin{minipage}[c]{0.24\textwidth}
    \raggedright  b) $\omega_c=745\,\mathrm{cm}^{-1}$
    \includegraphics[width=\textwidth]{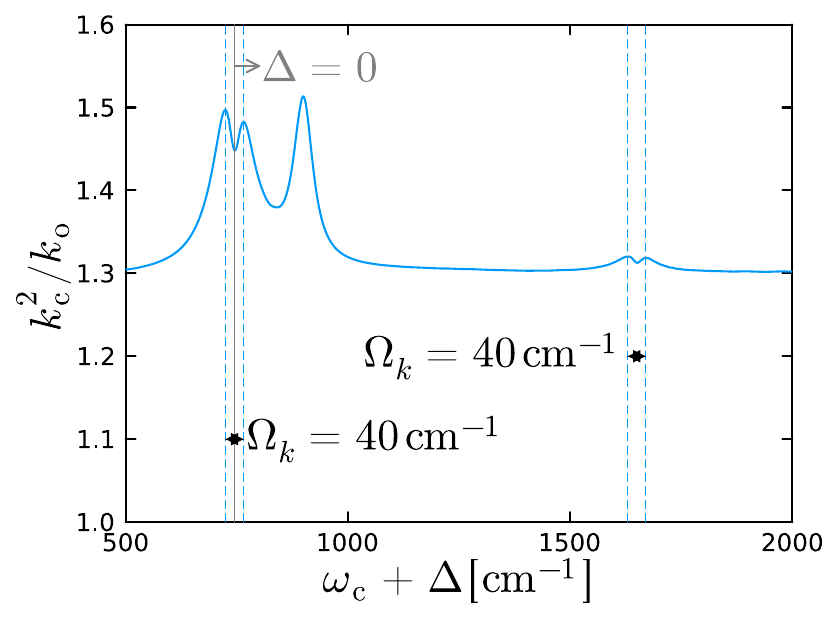}
  \end{minipage}
   \begin{minipage}[c]{0.24\textwidth}
    \raggedright  c) $\omega_c=900\,\mathrm{cm}^{-1}$
    \includegraphics[width=\textwidth]{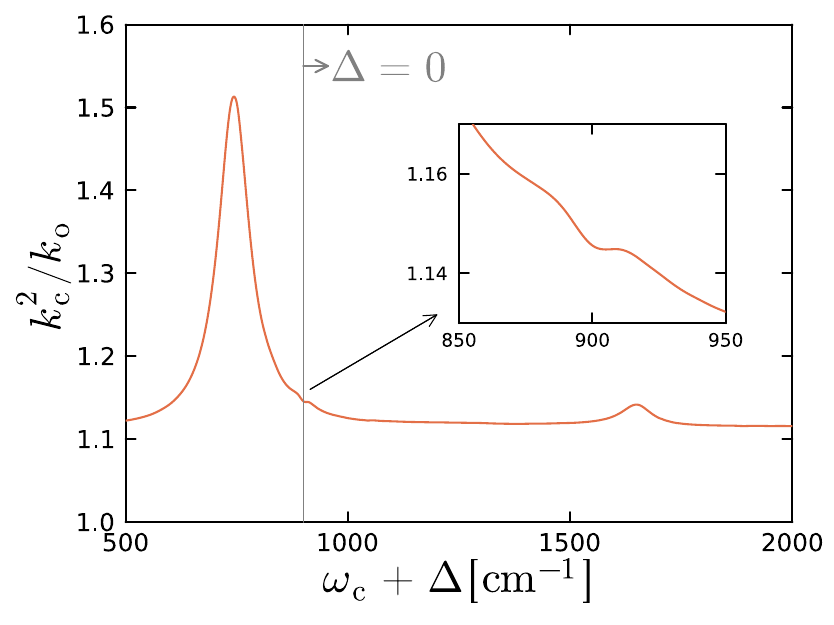}
  \end{minipage}
   \begin{minipage}[c]{0.24\textwidth}
    \raggedright  d) $\omega_c=1645\,\mathrm{cm}^{-1}$
    \includegraphics[width=\textwidth]{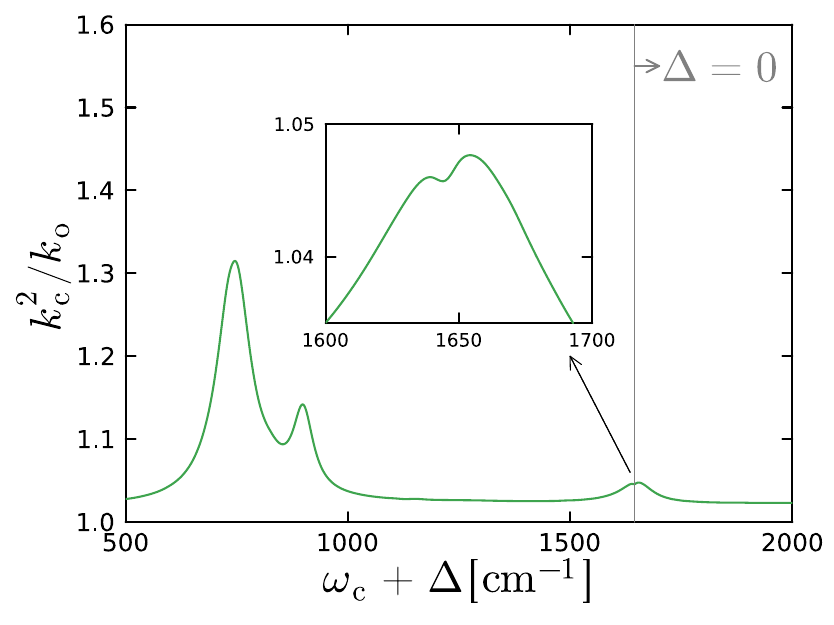}
  \end{minipage}
\caption{a) Rate modification factor ($k^1_{\mathrm c}/k_{\mathrm{o}}$) of Model II in a single-mode cavity as a function of the cavity frequency $\omega_{\mathrm{c}}$. b-c) Rate modification factor ($k^2_{\mathrm c}/k_{\mathrm{o}}$)  of Model II as a function of the neighboring cavity frequency $\omega'_{\mathrm{c}}$ in a two-mode cavity, where $\omega_{\mathrm c}$ is fixed in each panel, corresponding to three peaks in a). The light-matter coupling strength is set to $\eta_1=\eta_2=0.00125~$a.u.} \label{fig8:rates}
\end{figure*}

\begin{figure}
\centering
    \begin{minipage}[c]{0.45\textwidth}
    \includegraphics[width=\textwidth]{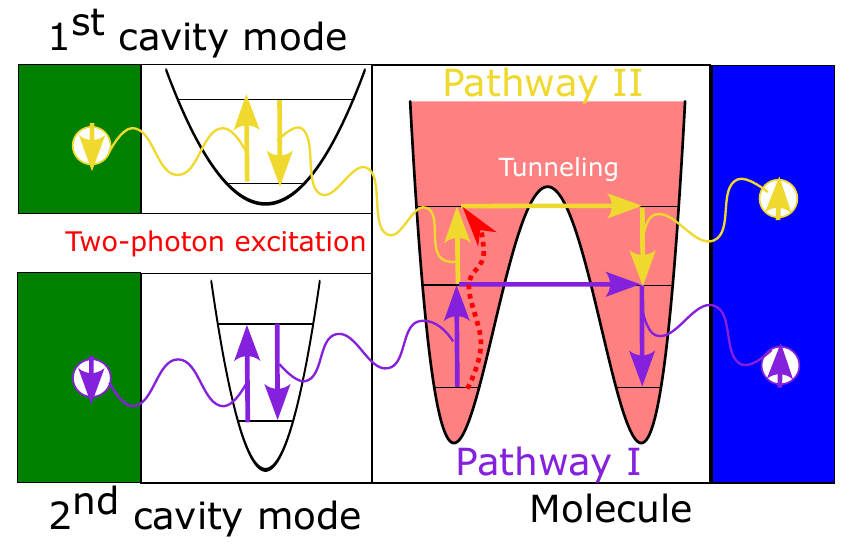}
  \end{minipage}
\caption{Schematic illustration of cavity-induced intramolecular reaction pathways in Model II. Pathway I starts with a cavity photon absorption (with energy $\omega_{\rm c}=900\,\mathrm{cm}^{-1}$) from the cavity bath, which is spontaneously emitted and captured by the molecule to induce vibrational excitation $|0^L\rangle\rightarrow |1^L\rangle$. Afterwards, the tunneling to the right well takes place, followed by a vibrational relaxation. The emitted quantized energy might dissipate into the solvent (or recaptured by the cavity mode, not shown). Pathway II is analogous other than that the cavity frequency is $\omega_{\rm c}=745\,\mathrm{cm}^{-1}$, which is resonant to and induces the vibrational transition $|1^L\rangle \rightarrow |2^L\rangle$. The red dotted curve illustrates a two-photon excitation step induced by the simultaneous coupling to two cavity modes under strong coupling conditions. } \label{fig9:schematic}
\end{figure}
\begin{figure}
\centering
    \begin{minipage}[c]{0.4\textwidth}
  \raggedright a) $\lambda_{\rm c}=100\,\mathrm{cm}^{-1}$
    \includegraphics[width=\textwidth]{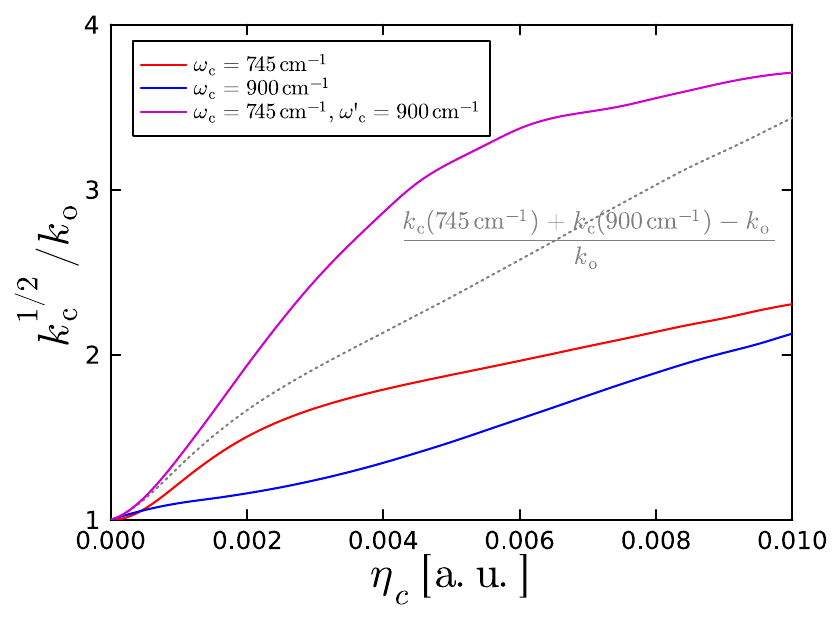}
  \end{minipage}
    \begin{minipage}[c]{0.4\textwidth}
  \raggedright b) $\eta_{\rm c}=0.00125\,{\rm a.u.}$ 
    \includegraphics[width=\textwidth]{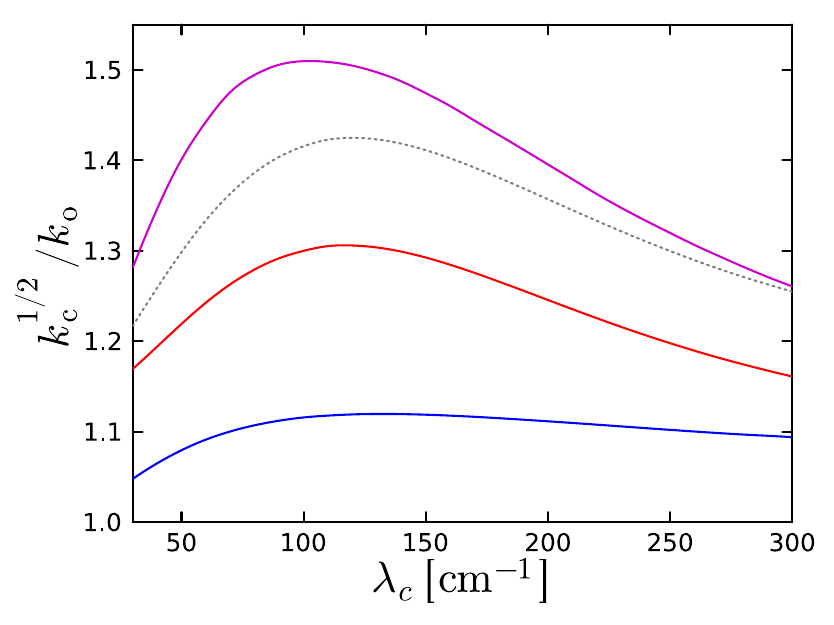}
  \end{minipage}
\caption{Comparison of the rate modification factors of Model II in single-mode cavities ($k^1_{\mathrm c}/k_{\mathrm{o}}$, with the red line for $\omega_{\mathrm{c}}=745\,\mathrm{cm}^{-1}$ and the blue line for $\omega_{\mathrm{c}}=900\,\mathrm{cm}^{-1}$) and in a two-mode cavity ($k^2_{\mathrm c}/k_{\mathrm{o}}$, with  $\omega_{\mathrm{c}}=745\,\mathrm{cm}^{-1}$ and $\omega'_{\mathrm{c}}=900\,\mathrm{cm}^{-1}$). In panel a), the results are shown as a function of the light-matter coupling strength $\eta_{\rm c}$, while $\lambda_{\rm c}=100\,\mathrm{cm}^{-1}$. In panel b), we keep $\eta_{\rm c}=0.00125$\au and vary the cavity loss strength $\lambda_{\rm c}$. As a reference, we also plot the additive ratio $r = (k^1_{\rm c}(745\,\mathrm{cm}^{-1}) + k^1_{\rm c}(900\,\mathrm{cm}^{-1}) - k_{\rm o})/k_{\rm o}$ as gray lines. } \label{fig10:sr}
\end{figure}

Next, we explore a distinct and practically relevant scenario in which multi-mode cavity effects can be strategically leveraged to enhance chemical reactivity. Unlike the previous case, here the cavity’s FSR is not necessarily tuned to match the energy scale of the single-mode Rabi splitting. Instead, the cavity is designed such that multiple resonant modes approximately match different vibrational transitions of an anharmonic reaction coordinate. In this configuration, the interplay between multi-mode coupling and molecular anharmonicity gives rise to a cooperative enhancement of the reaction rate—enabled not by mode near-degeneracy, but by the alignment of cavity modes with multiple intramolecular transitions that span the reaction pathway.

To this end, we retain all PES parameters from the previous model but increase the reduced mass associated with the reactive bond vibration to $M = 2$\au This adjustment lowers the vibrational level spacing and brings a greater number of energetically distinct transitions below the reaction barrier into play, as illustrated in \Fig{fig2:PES_absorption}~a2). While the detailed dynamics of this model in a single-mode cavity have been examined in our previous work,\cite{Ke_J.Chem.Phys._2024_p224704} we emphasize here that this system offers a compelling example where the complexity of reaction dynamics extends well beyond what is captured by linear spectroscopies.

The molecular absorption profile (see \Fig{fig2:PES_absorption}~b2)) is dominated by a sharp peak near $900\,\mathrm{cm}^{-1}$, corresponding to the ground-to-first excited state transition, $|0^{L/R}\rangle \leftrightarrow |1^{L/R}\rangle$, as indicated by the red double-headed arrows in both the left and right wells in \Fig{fig2:PES_absorption}~a2). Other dipole-allowed transitions are spectroscopically weak due to various limiting factors. For example, the small peak around $750\,\mathrm{cm}^{-1}$ (see the blue inset in \Fig{fig2:PES_absorption}~b2), corresponding to the transition $|1^{L/R}\rangle \leftrightarrow |2^{L/R}\rangle$--marked by the blue arrows in \Fig{fig2:PES_absorption}~a2), is suppressed due to exponentially vanishing thermal population in higher vibrational excited states. Here, the excited states $|2^{L/R}\rangle$ are defined as $|2^L\rangle=\frac{|v_{\rm m}=4\rangle +|v_{\rm m}=5\rangle}{\sqrt{2}}$ and $|2^R\rangle=\frac{|v_{\rm m}=4\rangle -|v_{\rm m}=5\rangle}{\sqrt{2}}$. Meanwhile, the overtone excitation $|0^{L/R}\rangle \leftrightarrow |2^{L/R}\rangle$, indicated by the green arrows in \Fig{fig2:PES_absorption}~a2), appears as a weak feature around $1670\,\mathrm{cm}^{-1}$ (see the green inset in \Fig{fig2:PES_absorption}~b2)), primarily due to the small transition dipole moment that limits its intensity.

In contrast to the absorption spectrum, the reaction dynamics reveal a markedly different picture regarding the contributions of these vibrational transitions. Notably, reaction rates are governed not solely by transition dipole strengths and thermal populations, but also critically shaped by dynamical factors such as tunneling efficiency and the timescale hierarchy between different reactive steps. In particular, tunneling near the barrier top proceeds significantly faster than in the vicinity of the potential well bottoms. As a result, transitions like $|1^{L/R}\rangle \leftrightarrow |2^{L/R}\rangle$, although spectroscopically weak, play a dominant role in enhancing the reaction rates inside the cavity. This occurs through their collaboration with the more rapid tunneling process near the barrier to enable a cavity-assisted intramolecular reaction pathway (see Pathway II in \Fig{fig9:schematic}). This mechanism leads to the strongest peak in the $k_{\rm c}^1/k_{\rm o}$ versus $\omega_{\rm c}$ profile under single-mode coupling, as shown in \Fig{fig8:rates}~a). By contrast, the spectroscopically dominant ground-to-first excited state transition, $|0^{L/R}\rangle \leftrightarrow |1^{L/R}\rangle$, contributes mainly to a slower, less efficient reaction channel (see Pathway I in \Fig{fig9:schematic}), yielding a weaker secondary peak in the single-mode rate modification profile. 

\Fig{fig8:rates}~b)–d) show the reaction rate ratios $k_{\rm c}^2 / k_{\rm o}$ in a two-mode cavity as a function of the second cavity mode frequency, $\omega'_{\rm c}$. Each panel corresponds to a fixed central cavity frequency, chosen from one of the three peak positions identified in \Fig{fig8:rates}~a). 

When the first cavity frequency is fixed at $\omega_{\rm c} = 745\,\mathrm{cm}^{-1}$, resonant with the excited state transition $|1^{L/R}\rangle \leftrightarrow |2^{L/R}\rangle$, the two-mode rate enhancement factor $k_{\rm c}^2/k_{\rm o}$ exhibits a much finer structure as a function of the second cavity frequency $\omega_{\rm c}'$. In this case, light-matter hybridization gives rise to the formation of polaritonic states $|{\rm LP}^{L/R}\rangle$  and $|{\rm UP}^{L/R}\rangle$ near the barrier top, resulting in a characteristic doublet in the rate enhancement profile when $\omega_{\rm c}'$ is tuned close to $\omega_{\rm c}$, as shown in \Fig{fig8:rates}~b). This splitting follows the same mechanism elucidated in the previous example. Intriguingly, another doublet emerges when $\omega'_{\rm c}\approx 1645\,\mathrm{cm}^{-1}$, near the overtone transition frequency.  This indicates that the dressing of molecular vibrational excited states near the barrier top by the first cavity mode also modifies the overtone transitions. As a result, transitions enabled by the second cavity mode from the hybridized vacuum state can access both polaritonic states near the barrier top on both the reactant and product sides. By contrast, tuning $\omega'_{\rm c}$ near $900\,\mathrm{cm}^{-1}$--matching the ground-to-first excited state transition--does not result in splitting, as this transition does not involve the hybridized polaritonic states. However, we notice that for $\omega'_{\rm c} \approx 900\,\mathrm{cm}^{-1}$, the ratio $k_{\rm c}^2/k_{\rm o}$ is further enhanced by a factor of approximately 0.2--roughly double the enhancement $(k_{\rm c}^1(900\,\mathrm{cm}^{-1})-k_{\rm o})/k_{\rm o}$ achieved in the corresponding single-mode cavity.
Here, $(k_{\rm c}^1(900\,\mathrm{cm}^{-1}))$ denotes the single-mode cavity reaction rate at $\omega_{\rm c}=900\,\mathrm{cm}^{-1}$. This observation points to a non-additive rate mechanism under multi-mode coupling, which will be expounded below.  

When $\omega_{\rm c} = 900\,\mathrm{cm}^{-1}$, a shallow splitting appears as $\omega'_{\rm c}$ is tuned in its vinicity (see the inset of \Fig{fig8:rates}~c)). The remarkable rate enhancement is only evident when a lower-frequency cavity mode $\omega'_{\rm c} \approx 750\,\mathrm{cm}^{-1}$, resonating with the $|1^{L/R}\rangle \leftrightarrow |2^{L/R}\rangle$ transition, is incorporated.
This again hints at the cooperative effect of two cavity modes when respectively matching distinct vibrational transitions. In contrast, when $\omega_{\rm c} = 1645\,\mathrm{cm}^{-1}$, no qualitative change is observed in the overall lineshape of \Fig{fig8:rates}~d) relative to the single-mode profile in \Fig{fig8:rates}~a). This is owing to the weak transition dipole strength of overtone excitation. The only notable features are a modest baseline increase and a small splitting in the high-frequency region near $\omega'_{\rm c}=1645\,\mathrm{cm}^{-1}$, as shown in the inset of \Fig{fig8:rates}~d).

To further investigate the non-additive rate enhancement effect discussed above, we consider a two-mode cavity with identical light-matter coupling strengths, $\eta_1=\eta_2=\eta_{\rm c}$ and fixed cavity frequencies $\omega_{\rm c} = 745\,\mathrm{cm}^{-1}$ and $\omega'_{\rm c} = 900\,\mathrm{cm}^{-1}$. The resulting reaction rate enhancement, $k_{\rm  c}^2 / k_{\rm  o}$, is shown in \Fig{fig10:sr}~a) as a function of the coupling strength $\eta_{\rm c}$. To quantify the degree of non-additivity, we also display the corresponding single-mode results and an additive estimate given by $r = (k^1_{\rm c}(745\,\mathrm{cm}^{-1}) + k^1_{\rm c}(900\,\mathrm{cm}^{-1}) - k_{\rm o})/k_{\rm o}$ (gray dotted line), where $k^1_{\rm c}(\omega_{\rm c})$ specifies the single-mode reaction rate for a cavity of frequency $\omega_{\rm c}$. In the weak coupling regime, the two-mode results closely follow the additive prediction, with $k_{\rm c}^2/k_{\rm o}$ nearly overlapping the line representing $r$. However, as $\eta_{\rm c}$ increases, the two-mode rate enhancement begins to exceed the additive estimate, signaling the onset of cooperative, non-additive behavior under strong coupling. 
This enhanced reactivity likely originates from a cavity-enabled two-photon intramolecular reaction pathway that emerges only when both vibrational transitions are strongly coupled to the respective cavity modes, as schematically illustrated by the red dotted curve in \Fig{fig9:schematic}. Specifically, cavity-induced excitation $|0^L\rangle \rightarrow |1^L\rangle$ transiently increases the population of the first vibrational excited state, thereby immediately facilitating the more efficient reaction Pathway II. This sequential activation, enabled by the cooperative interaction of both cavity modes, constitutes a key mechanism for the observed non-additive rate enhancement.

In our previous work,\cite{Ke_J.Chem.Phys._2025_p64702} we demonstrated that single-mode cavity-induced rate enhancement exhibits a stochastic resonance behavior with respect to $\lambda_{\rm c}$, the coupling strength between the cavity and its bath.  Specifically, the reaction rates under resonant cavity conditions display a non-monotonic dependence on the increasing external noise level $\lambda_{\rm c}$, characterized by a turnover. At low $\lambda_{\rm c}$, insufficient photon generation limits the initialization of cavity-induced reaction pathways, as illustrated in \Fig{fig9:schematic}. Conversely, at large $\lambda_{\rm c}$ meaning a fast energy exchange between the cavity mode and its bath, rapid energy dissipation depletes photon before they can be harnessed by the molecule. As a result, an intermediate noise level--corresponding to a moderate cavity-bath coupling--yields the optimal rate enhancement. This turnover behavior persists in the two-mode cavity setting as well. By fixing $\eta_{\rm c} = 0.00125\,\mathrm{a.u.}$, we compute the two-mode rate enhancement $k_{\rm c}^2 / k_{\rm o}$ as a function of $\lambda_{\rm c}$, as shown in \Fig{fig10:sr}~b). The results confirm that stochastic resonance remains a robust feature in multimode vibrational polaritonic dynamics, highlighting the essential interplay between cavity coherence and environmental noise in optimizing polariton-assisted chemical reactivity.

These results collectively underscore that strategic multi-mode cavity design--specifically, targeting its constituent mode frequencies to simultaneously match distinct vibrational transitions in the potential energy surface along the reaction coordinate--can yield cooperative, non-additive enhancements in chemical reactivity. 

\section{\label{sec:conclusion}Conclusion}
In this work, we have extended previous numerically exact quantum dynamical studies of chemical reactions in single-mode cavities to few-mode cavity environments. This extension aims to elucidate how the presence of multiple photonic modes—an inherent feature of most experimental cavity architectures—modifies the underlying reaction dynamics. To isolate and clarify the essential physical mechanisms, we confined our investigation to a single-molecule limit, providing foundational insight without the added complexity of collective effects.

Our simulations uncover two distinct scenarios where multi-mode effects play a critical role in reaction rate enhancement. The first occurs in low-finesse cavities, which can arise, for example, from an increased optical path length between two reflecting mirrors and are characterized by a reduced FSR. In such systems, multiple cavity modes may lie close in frequency to a molecular vibrational transition. When a molecular vibration—particularly one of those that significantly contribute to the rate-decisive step—strongly couples to a resonant high-order cavity mode, hybrid polaritonic states are generated. These states inherit both photonic and molecular character, thereby shifting the effective resonance conditions for neighboring modes. When these adjacent modes fall into resonance with the modified transitions, which is most likely when the FSR is comparable to the Rabi splitting, they can mediate additional reaction pathways, leading to a further rate enhancement beyond the single-mode case.

The second mechanism originates from the intrinsic anharmonicity of the molecular potential energy surface. In the strong coupling regime with multiple cavity modes, distinct vibrational transitions along the reaction coordinate—each associated with different transition energies—can simultaneously couple to individually resonant cavity modes. This enables a multi-photon absorption process in which each photon is drawn from a different mode, facilitating a sequential vibrational ladder climbing. As a result, the system is efficiently promoted from the vibrational ground state to highly excited states near the barrier top, where tunneling is significantly faster. This cascade excitation mechanism leads to a non-additive increase in the reaction rate.

Altogether, our results demonstrate that incorporating multi-mode cavity structure leads to qualitatively new reaction pathways and mechanisms for vibrational-polariton-assisted chemistry. We anticipate that this more realistic cavity description, going beyond the single-mode paradigm, will offer deeper insights into the design principles for polaritonic catalysis. Future work will focus on generalizing this framework to multi-mode, multi-molecule systems, where inter-molecular interactions and collective light–matter coupling introduce an even higher degree of complexity. Such systems pose substantial challenges for numerically exact quantum simulation, as the interaction network moves beyond the star-like topology tractable by tree tensor network methods. Addressing these challenges will require the development of more advanced quantum many-body algorithms and computational strategies.

\begin{acknowledgments}
The authors thank Prof. Jeremy Richardson for helpful discussions. Y. Ke thanks the Swiss National Science Foundation for the award of a research fellowship (Grant No. TMPFP2\_224947). 
\end{acknowledgments}

\section*{Supplementary information}
See the supplementary material for:  1) reaction rate modification profiles ($k_{\rm c}^1/k_{\rm o}$ versus $\omega_{\rm c}$) for varying $\lambda_{m}$, which demonstrating the broadening and merging of distinct vibrational transitions $|v_{\rm m}=0\rangle \leftrightarrow |v_{\rm m}=3\rangle$ and $|v_{\rm m}=1\rangle \leftrightarrow |v_{\rm m}=2\rangle$; 2) an analytical analysis of the reaction rate modifications in a two-mode cavity based on Fermi's Golden Rule rate theory, which fails to capture the correct dynamics; 3) the broadening effect induced by strong cavity dissipation at larger $\lambda_{\rm c}$, which blurs the hybridized light-matter states and washes out the splitting in the rate modification profile as a function of the FSR; 4) a comparison between the scenario where each cavity mode is coupled to its own cavity bath and the case where all cavity modes share a single bath; and 5) illustrative examples where the inclusion of additional cavity modes reduces the reaction rates. 

\section*{Data Availability Statement}
The data and code that support the findings of this work are available from the corresponding author upon reasonable request.

%

%

\pagebreak
\widetext
\clearpage
\begin{center}
\textbf{\large Supplementary information: Harnesing multi-mode optical structure for chemical reactivity}
\end{center}
\setcounter{equation}{0}
\setcounter{figure}{0}
\setcounter{table}{0}
\setcounter{page}{1}
\setcounter{section}{0}
\makeatletter
\renewcommand{\theequation}{S\arabic{equation}}
\renewcommand{\thefigure}{S\arabic{figure}}
\renewcommand{\bibnumfmt}[1]{[S#1]}
\renewcommand{\citenumfont}[1]{#1}
\renewcommand{\Sec}[1]{Sec.\,\ref{#1}}

\section{Broadening effect from the solvent bath}
\begin{figure}
\centering
    \begin{minipage}[c]{0.45\textwidth}
    \raggedright a) 
    \includegraphics[width=\textwidth]{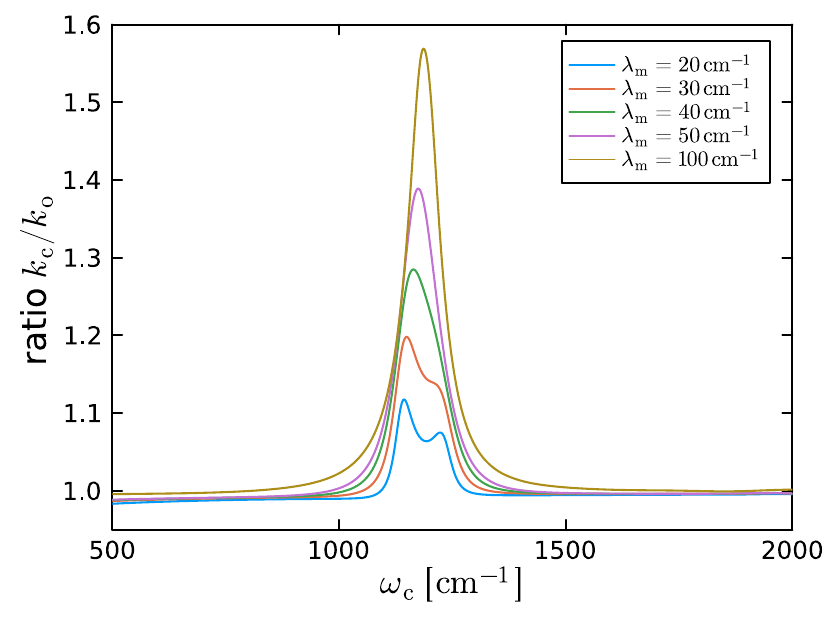}
  \end{minipage}
    \begin{minipage}[c]{0.45\textwidth}
        \raggedright b) 
    \includegraphics[width=\textwidth]{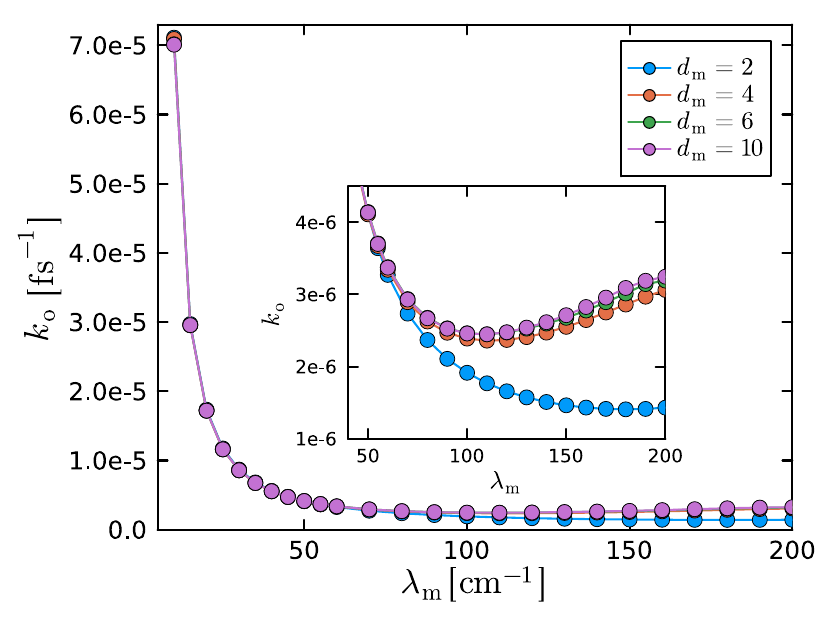}
  \end{minipage}
\caption{a) Rate modification profile $k_{\rm c}^1/k_{\rm o}$ in a single-mode cavity as a function of the cavity frequency $\omega_{\rm c}$ for different coupling strengths to the solvent bath. The light-matter coupling is $\eta_{\rm c}=0.00125$~a.u. b) Reaction rates outside the cavity $k_{\rm o}$ as a function of the molecule-solvent coupling strength $\lambda_{\rm m}$ for different $d_{\rm m}$ lowest vibrational states retained in the simulations.} \label{fig1:broadening}
\end{figure}

It is worth noting that, for the model system characterized by the potential energy surface shown in Fig. 2~a1) of the main text, two distinct dipole-allowed vibrational transitions are involved in the cavity-induced rate modification. The first transition occurs between the vibrational eigenstates $|v_m=0\rangle$ and $|v_m=3\rangle$, with a transitional energy of $\delta E_{0\leftrightarrow3}=1238\,\mathrm{cm}^{-1}$. The second transition takes place between the neighboring states $|v_m=1\rangle$ and $|v_m=2\rangle$, with a slightly lower energy gap of $\delta E_{1\leftrightarrow 2}=1140\,\mathrm{cm}^{-1}$. When system-bath coupling is weak--specifically, when both the direct molecule-solvent interaction $\lambda_{\rm m}$ and the indirect broadening effects mediated through the cavity mode ($\eta_{\rm c}$) and cavity baths are minimal--these two vibrational transitions remain energetically well-resolved. As a result, they manifest as two peaks in the cavity-modified reaction rate profiles as a function of the cavity frequency $\omega_{\rm c}$. This is illustrated in \Fig{fig1:broadening} a), which shows the single-mode reaction rate ratio $k^1_{\rm c}/k_{\rm o}$ as a function of the cavity frequency $\omega_{\rm c}$ for various values of the molecule–solvent coupling strength $\lambda_m$. In the weak-damping limit (small $\lambda_m$), two distinct peaks appear,  centered at the cavity frequencies resonant with the two molecular vibrational transitions. This indicates that the cavity can selectively couple to either transition, modifying the reaction rate through different dynamical pathways.

As $\lambda_m$ increases, the reaction rate in the absence of the cavity (i.e., $k_{\rm o}$) initially decreases, as shown in \Fig{fig1:broadening} b). This behavior arises from suppressed tunneling. To examine this in more detail, we present the results obtained with different $d_{\rm m}$, the number of the lowest vibrational eigenstates retained in the simulations. When $d_{\rm m}=2$, the reaction proceeds purely via the tunneling between two nearly degenerate vibrational ground states at the bottom of two wells. In this case, the rates $k_{\rm o}(d_{\rm m}=2)$ decrease monotonically with increasing $\lambda_m$ over the studied range, reflecting that strong coupling to the solvent inhibits pure ground-state tunneling. When $\lambda_m>50\,\mathrm{cm}^{-1}$, convergence requires $d_{\rm m}>2$, which signals the involvement of higher vibrational states. In this regime, solvent-induced vibrational heating and cooling, together with the faster tunneling through higher vibrationally excited states, establish an additional cotunneling pathway. This pathway becomes increasingly important as the molecule-solvent coupling strengthens.  Consequently, for $\lambda_m>100\,\mathrm{cm}^{-1}$, the contribution of this cotunneling pathway outweighs the suppression of ground-state tunneling, leading to an overall increase in the converged reaction rates with growing $\lambda_{\rm m}$. 

Inside the cavity, the enhancement of the reaction rates becomes more pronounced with increasing $\lambda_{\rm m}$. This behavior can be attributed to the faster energy exchange between the molecule and the solvent bath within the cavity-induced intramolecular reaction pathway, as illustrated in Fig. 5~b) of the main text. With sufficiently strong $\lambda_{\rm m}$, the spectral broadening caused by increasing $\lambda_{m}$ eventually leads to the coalescence of two initially separated peaks into a single broadened feature. In this regime, the cavity no longer resolves the two transitions separately. Instead, the vibrational transitions are effectively coarse-grained, and the reaction dynamics can be better delineated in terms of the degenerate transitions between the localized vibrational ground states $|0^{L/R}\rangle$ and first excited states $|1^{L/R}\rangle$ within each well, as shown in Fig. 2~a) of the main text. 

Importantly, we highlight the conceptual distinction between the peak splitting observed in \Fig{fig1:broadening} for the single-mode rate modification profile under weak damping and the energy splitting that emerges when the second cavity mode is introduced. In the former case, the multiple peaks arise from the intrinsic molecular vibrational structure--i.e., the presence of multiple dipole-allowed vibrational transitions with distinct energies. This is purely a molecular feature and does not require strong light-matter coupling.

In contrast, the splitting observed in the two-mode cavity setup--evident in the profile of $k^2_{\rm c}/k_{\rm o}$ as a function of the free spectral range $\Delta$--originates from the light-matter hybridization. That is, the resonant interaction between a molecular transition and a cavity mode leads to the formation of polaritonic states (e.g., upper and lower polaritons), both of which have partial molecular excited state character and their energetic separation scales with the light-matter coupling strength.  As a consequence, under sufficiently strong coupling conditions, the polaritonic energy splitting can exceed the original energy difference between the two distinct molecular vibrational transitions ($|\delta E_{0\leftrightarrow 3}-\delta E_{1\leftrightarrow 2}|=98\,\mathrm{cm}^{-1}$), which is the case observed in Fig. 6~b) of the main text. This behavior underscores a fundamentally emergent feature in the polaritonic chemistry: the appearance of new reactive channels in a multi-mode cavity must explicitly take into account the light-matter hybridization, and cannot be understood or simulated by simply shoving the quantized cavity modes into the environment, which might underlies the failure of a naive implementation of Fermi's Golden Rule rate theory, as detailed in the following section.

\begin{figure}
\centering
\begin{minipage}[c]{0.45\textwidth}
\raggedright a1) single-mode, HEOM+TTNS
    \includegraphics[width=\textwidth]{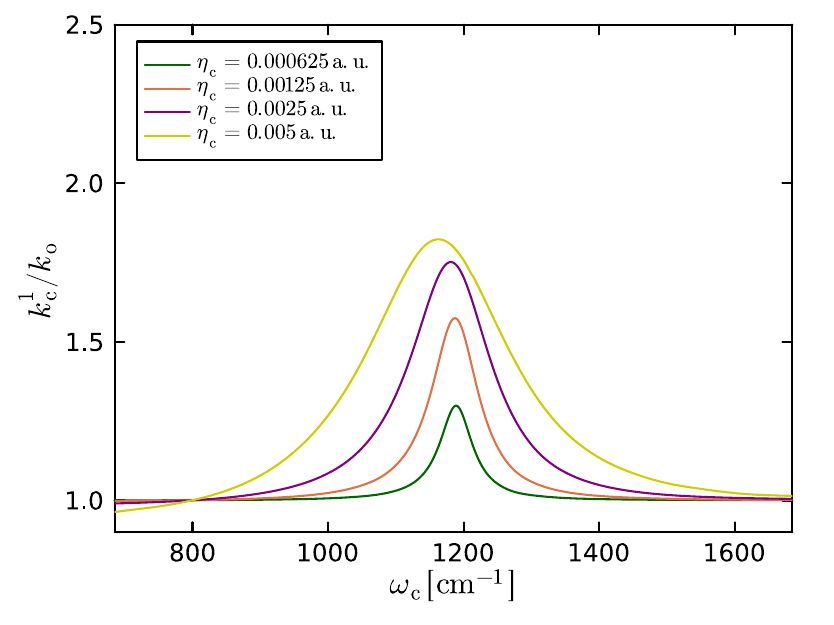}
  \end{minipage}
    \begin{minipage}[c]{0.45\textwidth}
    \raggedright a2) two-mode, HEOM+TTNS
    \includegraphics[width=\textwidth]{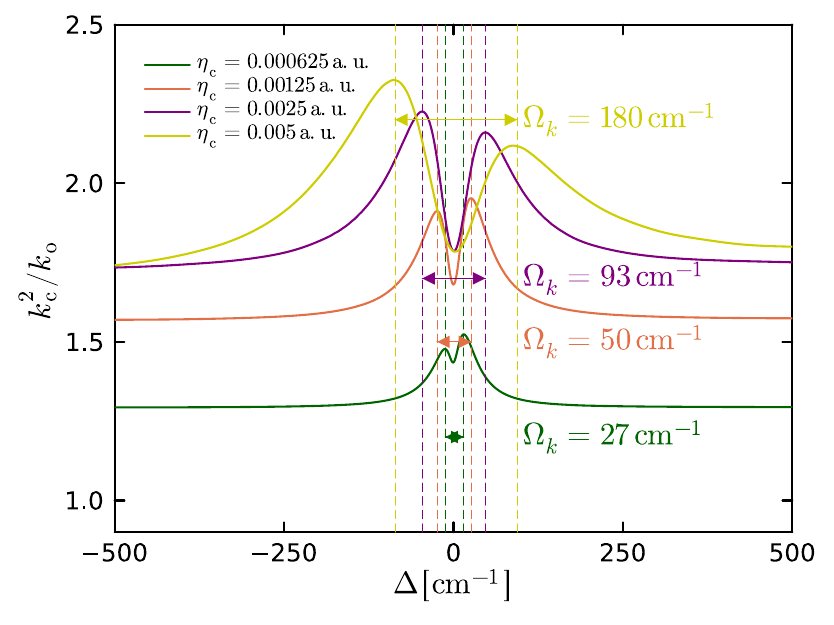}
  \end{minipage}
     \begin{minipage}[c]{0.45\textwidth}
  \raggedright b1) single-mode, FGR
    \includegraphics[width=\textwidth]{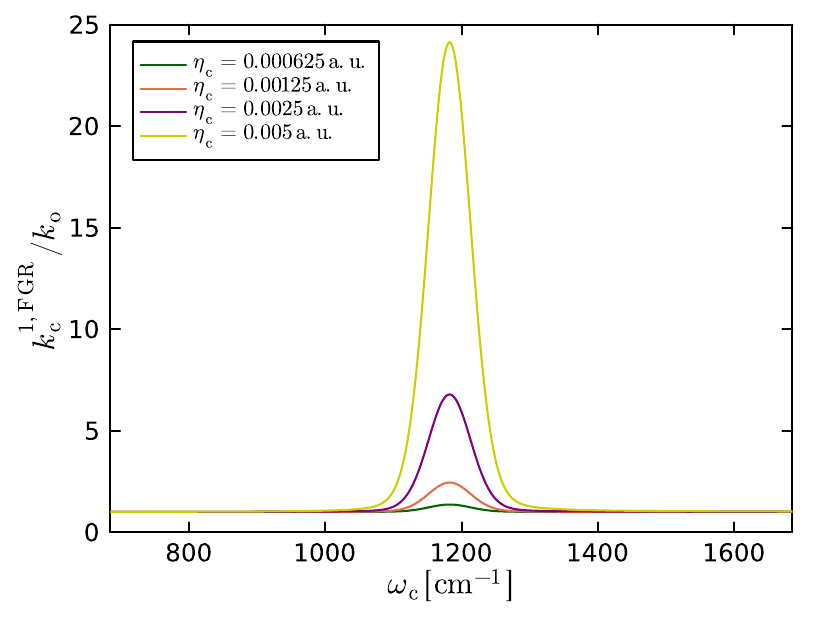}
  \end{minipage}
    \begin{minipage}[c]{0.45\textwidth}
    \raggedright b2) two-mode,  FGR
    \includegraphics[width=\textwidth]{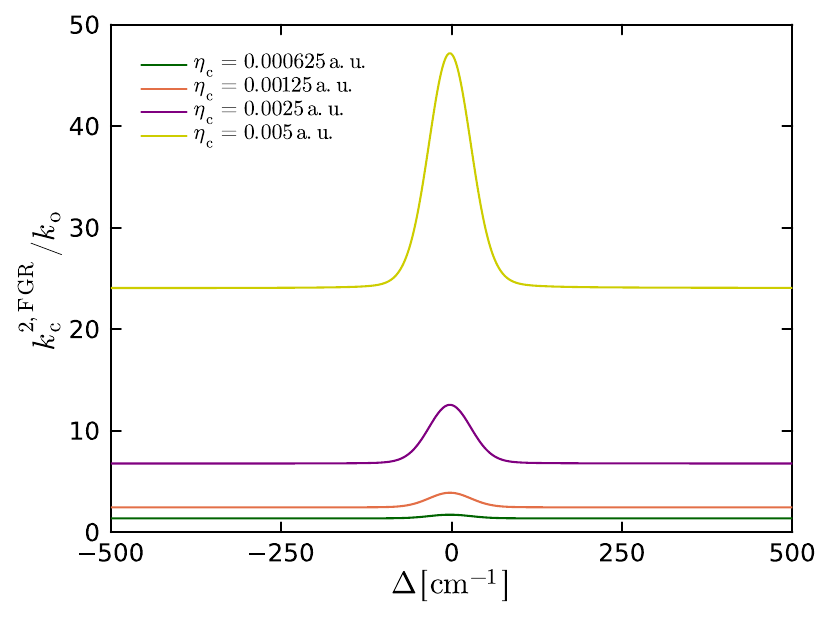}
  \end{minipage}
       \begin{minipage}[c]{0.45\textwidth}
  \raggedright c1) single-mode, modified FGR
    \includegraphics[width=\textwidth]{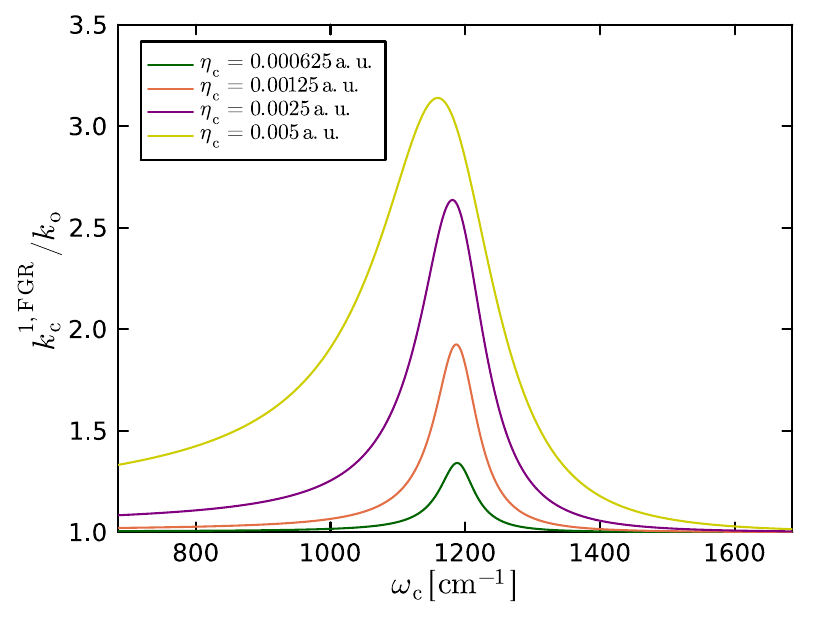}
  \end{minipage}
    \begin{minipage}[c]{0.45\textwidth}
    \raggedright c2) two-mode,  modified FGR
    \includegraphics[width=\textwidth]{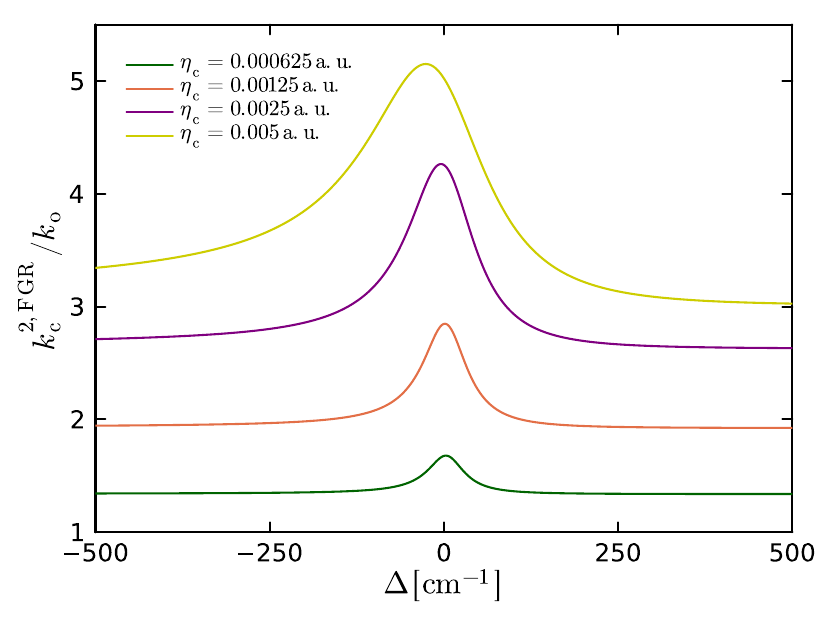}
  \end{minipage}
\caption{Rate modification profile $k_{\rm c}^1/k_{\rm o}$ for Model I in a single-mode cavity as a function of the cavity frequency $\omega_{\rm c}$ for different light-matter coupling strength $\eta_c$, obtained using the HEOM+TTNS approach in panel a1), the FGR rate theory in panel b1), and the modified FGR rate theory in panel c1).  Rate modification profile ($k_{\rm c}^2/k_{\rm o}$) for the same molecule in a two-mode cavity as a function of the free spectral range $\Delta$, obtained using the HEOM+TTNS approach in panel a2), the FGR rate theory in panel b2), and the modified FGR theory in panel c2).   Here, we assume the coupling strengths of two cavity modes to the molecule are the same: $\eta_1=\eta_2=\eta_c$. The results are displayed for different values of $\eta_{\rm c}$, while the central cavity frequency is fixed at $\omega_{\rm c}=1185\,\mathrm{cm}^{-1}$, and $\lambda_{\rm c}=100\,\mathrm{cm}^{-1}$. For the modified FGR, the fitting parameter is $\alpha = 0.5$.} \label{fig2:HEOM_FGR}
\end{figure}

\section{Fermi's Golden Rule rate calculation for multi-mode optical cavity induced reaction rate modification}

Motivated by the promising findings in Ref.~\onlinecite{Ying_2023_JCP_p84104}, which utilized Fermi’s Golden Rule (FGR) to qualitatively describe reaction rate modifications induced by a single-mode cavity and works for the symmetric double-well potential with $M=1$\au (see Fig.~2a) of the main text), we investigate whether this formalism can be naturally extended to multi-mode cavity scenarios.

In Ref.~\onlinecite{Ying_2023_JCP_p84104}, a critical step is the recasting of the total Hamiltonian to isolate the molecular degree of freedom as the system, while subsuming the cavity mode into the cavity bath to form an effective bath. The total Hamiltonian with the cavity modes and cavity baths given explicitly reads
\begin{align}\label{starting-point}
H =& \frac{p_{\mathrm{m}}^2}{2M}+ U(x_\mathrm{m}) + H_E^{\rm m}\nonumber+ \sum\limits_i \omega_{i} \eta^2_ix_\mathrm{m}^2 \\
&+ \sum\limits_i \left[\sqrt{2\omega_i^3}\eta_ix_\mathrm{m}x_i + \frac{p_i^2}{2} 
+ \frac{\omega_i^2}{2} x_i^2  +  \sum\limits_{k}\frac{P^2_{ik}}{2} + \frac{\omega^2_{i k}}{2}\left( Q_{i k} + \frac{g_{i k}x_i}{\omega_{i k}^2}\right)^2\right],
\end{align}
where $x_\mathrm{m}$ denotes the molecular reaction coordinate, $x_i$ are the cavity mode coordinates, and $Q_{ik}$ are the cavity bath coordinates coupled to the cavity mode $i$.

To recast the problem in a form that permits an effective spectral density description, a normal mode transformation is applied,\cite{leggett_quantum_1984,garg_effect_1985} to eliminate the explicit appearance of the cavity mode coordinates $x_i$, yielding:
\begin{align}
\label{normal-mode-transformed}
H &= \frac{p_{\mathrm{m}}^2}{2M} + U(x_{\mathrm{m}}) + H_E^{\rm m}\nonumber+ \sum\limits_i \omega_{i} \eta^2_ix_\mathrm{m}^2\nonumber\\
&\quad + \sum_i \left[
\sum_k \tilde{g}_{\t i \t k} \tilde{Q}_{i k}x_{\mathrm{m}}
+ \sum_k \left( \frac{\tilde{P}_{i k}^2}{2} + \frac{\tilde{\omega}_{i k}^2 }{2}\tilde{Q}_{i k}^2\right)
\right],
\end{align}
where $\tilde{Q}_{ik}$, $\tilde{P}_{ik}$, $\tilde{\omega}_{ik}$, and $\tilde{g}_{ik}$ represent the transformed effective bath coordinates, momenta, frequencies, and coupling constants to the system dipole operator $\mu(x_\mathrm{m})=x_\mathrm{m}$. This transformation allows the influence of the cavity and its associated bath to be encoded entirely in an effective spectral density:
\begin{equation}
    J_{\t{eff}}(\omega) = \frac{\pi}{2} \sum_i\sum_k \frac{\tilde{g}^2_{ik}}{\tilde{\omega}_{ik}}\delta (\omega-\tilde{\omega}_{ik}).
\end{equation}

In what follows, we follow the derivations in Ref.~\onlinecite{Ying_2023_JCP_p84104} to obtain this effective spectral density function for a multi-mode cavity. 

The classical equations of motion from the transformed Hamiltonian in \Eq{normal-mode-transformed} is given by
\begin{subequations}
\label{classicalEOM_normalized}
    \begin{align}
    \label{classicalEOM_normalized_1}
     M \ddot{x}_{\rm m} &= - V'(x_{\rm m}) - \sum\limits_i \left[2\omega_{i}\eta^2_ix_\mathrm{m}+\sum_k \tilde{g}_{\t i \t k} \tilde{Q}_{i k}\right], \\
    \label{classicalEOM_normalized_2}
        \ddot{Q}_{ik} &= -\tilde{\omega}_{ik}^2\tilde{Q}_{ik} -\tilde{g}_{ik}x_{\rm m}.
    \end{align}
\end{subequations}
where $V'(x_{\rm m}) =\frac{\partial U(x_{\rm m})}{\partial x_{\rm m}}+\frac{\partial H_{E}^{\rm m}}{\partial x_{\rm m}}$.
The Fourier transform of \Eq{classicalEOM_normalized} yields
\begin{subequations}
\label{classicalEOM_normalized_FT}
    \begin{align}        
    \label{classicalEOM_normalized_FT_1}
     -M \omega^2{x}_{\rm m}(\omega) &= - V'_{\omega}(x_{\rm m})  - \sum\limits_i \left[2\omega_{i}\eta^2_ix_\mathrm{m}(\omega)+\sum_k \tilde{g}_{\t i \t k} \tilde{Q}_{i k}(\omega)\right], \\
    \label{classicalEOM_normalized_FT_2}
       -\omega^2 \tilde{Q}_{ik}(\omega) &= -\tilde{\omega}_{ik}^2\tilde{Q}_{ik}(\omega) -\tilde{g}_{ik}x_{\rm m}(\omega), 
    \end{align}
\end{subequations}
where $V'_{\omega}(x_{\rm m})$ is the Fourier transform of $V'(x_{\rm m})$.
Reformulating \Eq{classicalEOM_normalized_FT_2} as $\tilde{Q}_{ik}(\omega)=\frac{-\tilde{g}_{ik}x_{\rm m}(\omega)}{\tilde{\omega}_{ik}^2-\omega^2}$ and substituting back in into \Eq{classicalEOM_normalized_FT_1}, we arrive at a susceptibility kernel:
\begin{equation}
K(\omega)
:= - M \omega^2  
+  \sum\limits_i2\omega_i\eta_i^2 
-  \sum\limits_i\sum_k \frac{\tilde{g}_{ik}^2}{\tilde{\omega}_{ik}^2
- \omega^2}     
= -V'_{\omega}(x_{\rm m})/x_m(\omega).
\end{equation}

Recognizing the bath sum as an integral over the spectral density:
\begin{equation}
    \sum_i\sum_k \frac{\tilde{g}_{ik}^2}{\tilde{\omega}_{ik}^2
- \omega^2} = \int_0^{\infty} \sum_i\sum_k  \frac{\tilde{g}_{ik}^2}{\tilde{\omega}_{ik}}\delta(s-\tilde{\omega}_{ik}) \frac{s}{s^2
- \omega^2} \mathrm{d}s = \int_0^{\infty} \sum_i\sum_k  J_{\t{eff}}(s) \frac{s}{s^2
- \omega^2} \mathrm{d}s, 
\end{equation}
and introducing a vanishing imaginary part ($-i\epsilon$) to $\omega$ for regularization, the imaginary component of the memory kernel $K(\omega)$ yields the effective spectral density:\cite{weiss_quantum_2022}
\begin{align}
\label{ImK}
        \lim\limits_{\epsilon\to 0^+} \Im  K(\omega-i\epsilon) 
    =& 
    \lim\limits_{\epsilon\to 0^+} \Im\int\limits_0^\infty -\frac{2}{\pi
    }\frac{J_{\mathrm{eff}}(s)s}{s^2-\omega^2 + i\epsilon}\t ds 
    =
    \int\limits_0^\infty 2J_{\mathrm{eff}}(s)s\delta(s^2-\omega^2)\t ds
    =
       \int\limits_0^\infty J_{\mathrm{eff}}(s)\frac{2s}{|2s|}\delta(s-\omega)\t ds \nonumber\\
       = &J_{\mathrm{eff}}(\omega).
\end{align}
Here, we have used the distributional identity 
\begin{equation}
\label{identity}
\lim\limits_{\epsilon\to0^+} \frac{1}{x-i\epsilon} = PV \frac{1}{x}+ i\pi\delta(x),
\end{equation}
where $PV$ indicates a principal value integral.

In parallel, we can derive the same quantity, the memory kernel $K(\omega)$ from the original Hamiltonian in \Eq{starting-point}. The classical equations of motion in terms of the original Hamiltonian give rise to
\begin{subequations}
\label{classicalEOM}
    \begin{align}
    M\ddot{x}_{\rm m}  
    &=  -V'(x_{\rm m}) 
    - \sum_i\left[2\omega_i\eta^2_i x_{\rm m}  
    + \omega_i^2\zeta_ix_i\right],
    \label{classicalEOM_1} \\
    \ddot{x}_{i} 
    &= -\sqrt{2\omega_i^3}\eta_i x_{\rm m} 
    - \omega_i^2 x_i - \sum_k g_{ik}\left(Q_{ik}
    +\frac{g_{ik}x_i}{\omega_{ik}^2}\right), 
    \label{classicalEOM_2}\\
     \ddot{Q}_{ik} 
    &= -\omega_{ik}^2\left(Q_{ik} 
    + \frac{g_{ik}x_i}{\omega_{ik}^2}\right).
    \label{classicalEOM_3}
\end{align}
\end{subequations}
Again, applying the Fourier transform to \Eq{classicalEOM}, we get
\begin{subequations}
\label{classicalEOM_FT}
    \begin{align}
    -M\omega^2 x_{\rm m} (\omega) 
    &=  - V'_\omega(x_{\rm m})
    - \sum_i\left[2\omega_{i}\eta^2_i x_{\rm m}(\omega)  
    + \sqrt{2\omega_i^3}\eta_ix_i(\omega)\right],
    \label{classicalEOM_FT_1} \\
    -\omega^2x_i(\omega) 
    &= -\sqrt{2\omega_i^3}\eta_i x_{\rm m}(\omega) 
    - \omega_i^2 x_i(\omega) - \sum_k g_{ik}\left(Q_{ik}(\omega)
    +\frac{g_{ik}x_i(\omega)}{\omega_{ik}^2}\right) ,
    \label{classicalEOM_FT_2}\\
    -\omega^2 Q_{ik}(\omega) 
    &= -\omega_{ik}^2\left(Q_{ik}(\omega) 
    + \frac{g_{ik}x_i(\omega)}{\omega_{ik}^2}\right).
    \label{classicalEOM_FT_3}
\end{align}
\end{subequations}
First, we reformulate \Eq{classicalEOM_FT_3} as $Q_{ik}(\omega)=\frac{-g_{ik}}{\omega_{ik}^2-\omega^2}x_i(\omega)$, and substitute $Q_{ik}(\omega)$ back in \Eq{classicalEOM_FT_2} to obtain
\begin{equation}
\label{xi}
\left( \omega_i^2-\omega^2
    -\sum\limits_k\frac{\omega^2g_{ik}^2}{\omega_{ik}^2(\omega_{ik}^2-\omega^2)}\right)x_i(\omega) 
    = -\sqrt{2\omega_i^3}\eta_ix_{\rm m}(\omega).
\end{equation}
Next, we introduce 
\begin{equation}
\label{Li}
    L_i(\omega) 
    := -\omega^2\left(1 + \sum_k \frac{g_{ik}^2}{\omega_{ik}^2(\omega_{ik}^2-\omega^2)}\right) 
    = 
    -\omega^2\left(1 + \frac{2}{\pi}\int\limits_0^\infty \frac{J_i(s)}{s(s^2-\omega^2)}\mathrm{d}s\right).
\end{equation}
where $J_i(s)=\frac{\pi}{2}\sum_k\frac{g_{ik}^2}{\omega_{ik}}(s-\omega_{ik})$ is the spectral density function for the original cavity bath $i$, so as to simplify 
\begin{equation}
  x_i(\omega) = -\frac{\sqrt{2\omega_i^3}\eta_i}{\omega_i^2+L_i(\omega)}   x_{\rm m}(\omega).
\end{equation}
Plugging this expression into \Eq{classicalEOM_FT_1}, the memory kernel $K(\omega)$ becomes
\begin{equation}
\label{KM}
    K(\omega)
    = -M\omega^2
    + \sum_i    \frac{2\omega_i\eta_i^2L_i(\omega)} {\omega_i^2+L_i(\omega)}
    =  -V'_{\omega}(x_{\rm m})/x_{\rm m}(\omega).
\end{equation}
Hence, taking the imaginary part of $K(\omega-i\epsilon)$, as per \Eq{ImK},  we recover the effective spectral density for a multi-mode cavity as:
\begin{align}
    J_{\mathrm{eff}}(\omega)
    = \lim\limits_{\epsilon\to 0^+}\Im K(\omega-i\epsilon) 
    = \sum\limits_i\frac{2\omega_i^3\eta_i^2J_i
    (\omega)}{\left(\omega_i^2-\omega^2 +  \tilde{R}_i(\omega)\right)^2+J_i(\omega)^2},
\end{align}
with 
\begin{equation}
    \tilde{R}_i(\omega)
    := \lim\limits_{\epsilon\to 0^+} \Re \left[ -\frac{2\omega^2}{\pi}\int\limits_0^\infty \frac{J_i(s)}{s\left(s^2-(\omega-i\epsilon)^2\right)}\t d s \right]
    = \frac{2\omega^2}{\pi}PV\int\limits_0^\infty\frac{J_i(s)}{s(\omega^2-s^2)}\t ds.
\end{equation}
Here, we used the identity in \Eq{identity} again to evaluate the integral in $L_i(\omega)$. For the Debye-Lorentzian spectral density function as given in Eq. (10) in the main text, this principal value integral can be evaluated analytically as $\tilde{R}_i(\omega) = \frac{\omega J_i(\omega)}{\Omega_c}$. Thus, in the multi-mode case, the overall effective spectral density is just a simple sum of each mode and the associated bath's contribution.

Adopting the same assumption as in Ref.~\onlinecite{Ying_2023_JCP_p84104}—that the vibrational transition between the ground state $|0^L\rangle$ and the first excited state $|1^L\rangle$ in the reactant region is rate-limiting--although this has been proven not universally true in our previous work\cite{ke2025non}--one can express the total reaction rate inside the cavity as:
\begin{equation}
    k^{\t{FGR}}_{\rm c} = k_{\rm o} + \alpha k_{\t{VSC}},
\end{equation}
where $k_0$ is the outside-cavity rate, $k_{\mathrm{VSC}}$ is the cavity-modified vibrational transition rate, and $\alpha$ is a fitting parameter obtained by comparison of $k^{\t{FGR}}_{\rm c}$ to the exact  results $k_{\rm c}$.

The expression for the cavity-modified vibrational transition rate $k_{\t{VSC}}$ is given by
\begin{equation}
\label{FGR_rate}
    k_{\t{VSC}} = 2|\mu_{01}^L|^2\int\limits_0^\infty J_{\t{eff}}(\omega)n(\omega)G(\omega)\t d\omega,
\end{equation}\label{fgr-expression}
where $\mu_{01}^L=\langle 1^L|\mu(x_{\rm m})|0^L\rangle)$ is the transition dipole moment,  $n(\omega)= \frac{1}{e^{\beta\omega}-1}$ is the Bose-Einstein distribution function, and $G(\omega_0, \sigma^2)$ is a normal distribution, which accounts for Gaussian broadening effects from the coupling to the solvent bath. The expectation value $\omega_0$ is the energy gap between states $|0^L\rangle$ and $1^L\rangle$, which is explicitly given by $\omega_0=\frac{\delta E_{0\leftrightarrow 3}+\delta E_{1\leftrightarrow 2}}{2}$. The variance is given by
\begin{equation}
    \sigma^2
    = \epsilon_z^2\sum\limits_k \frac{g_{\rm m k}^2}{2\omega_{\rm m k}}\coth\left(\frac{\beta\omega_{\rm m k}}{2}\right)
    = \epsilon_z^2\frac{1}{\pi}\int\limits_0^\infty J_{\rm m}(\omega)\coth\left(\frac{\beta\omega}{2}\right)\t d\omega,
\end{equation}
where $\epsilon_z=\langle 1^L| \mu(x_{\rm m})|1^L\rangle-\langle 0^L| \mu(x_{\rm m})|0^L\rangle$.

We verified our implementation of \Eq{FGR_rate} by reproducing results from Ref.~\onlinecite{Ying_2023_JCP_p84104}. \Fig{fig2:HEOM_FGR} a1) and b1) display the rate modification profile $k_{\rm c}^1/k_{\rm o}$ for Model I in a single-mode cavity as a function of the cavity frequency $\omega_{\rm c}$ at different light-matter coupling strengths $\eta_c$, obtained using the HEOM+TTNS approach and the FGR rate theory, respectively. For a single-mode cavity in the weak light-matter coupling regime and under the Markovian limit, using $\alpha = 0.4$, as suggested by Ying and Huo,\cite{Ying_2023_JCP_p84104}, the FGR rate well reproduces the qualitative frequency-dependent trends. However, this analytical rate formula significantly overestimates enhancements in the strong-coupling regime and fails to capture the red-shift and peak broadening observed in exact HEOM-TTNS results. The reason is that, in evaluating $k_{\rm VSC}$ via \Eq{FGR_rate}, it is implicitly assumed that the energy exchange between the cavity mode and the molecular vibrational transition $|0^L\rangle\rightarrow |1^L\rangle$ is always the rate-limiting step. This assumption breaks down in the strong light-matter coupling regime, where the bottleneck instead shifts to the photon exchange between the cavity mode and its bath. 

To remedy this, Ying and Huo proposed a modified FGR rate theory in Ref.~\onlinecite{Ying_2024_CM_p110} that yields improved results in the weak-loss regime. The modified rate expression is given by
\begin{equation}
    k^{\t{FGR}}_{\rm c} = k_{\rm o} + \alpha \tilde{k}_{\t{VSC}},
\end{equation}
where 
\begin{equation}
\label{modified_FGR_rate}
    \tilde{k}_{\t{VSC}} = \frac{2\pi\eta_{\rm c}^2\omega_{\rm c}^2|\mu_{01}^L|^2 \mathcal{A}(\omega_{\rm c}-\omega_0)n(\omega_{\rm c}) }{1+2\pi\eta_{\rm c}^2\omega_{\rm c}^2|\mu_{01}^L|^2 \mathcal{A}(\omega_{\rm c}-\omega_0)\tau_{\rm c}}
  \end{equation}
with the optical lineshape function 
\begin{equation}
    \mathcal{A}(\omega-\omega_0) =\frac{1}{\pi} \frac{\sigma}{(\omega-\omega_0)^2+\sigma^2}.
\end{equation}
Here $\tau_{\rm c}$ is the cavity lifetime, defined as $\tau_{\rm c}=\frac{\omega_{\rm c} (1-e^{-\beta \omega_{\rm c}})}{ J_{\rm c} (\omega_{\rm c})}$. Likewise, a straightforward extension to the two-mode case gives
\begin{equation}
\label{modified_FGR_rate_twomode}
    \tilde{k}_{\t{VSC}} = \sum_i\frac{2\pi\eta_{\rm i}^2\omega_{\rm i}^2|\mu_{01}^L|^2 \mathcal{A}(\omega_{\rm i}-\omega_0)n(\omega_{\rm i}) }{1+2\pi\eta_{\rm i}^2\omega_{\rm i}^2|\mu_{01}^L|^2 \mathcal{A}(\omega_{\rm i}-\omega_0)\tau_{\rm i}}.
  \end{equation}

\begin{figure}
\centering
    \begin{minipage}[c]{0.45\textwidth} 
      \raggedright a) $\lambda_{\rm c}=50\, \mathrm{cm}^{-1}$
    \includegraphics[width=\textwidth]{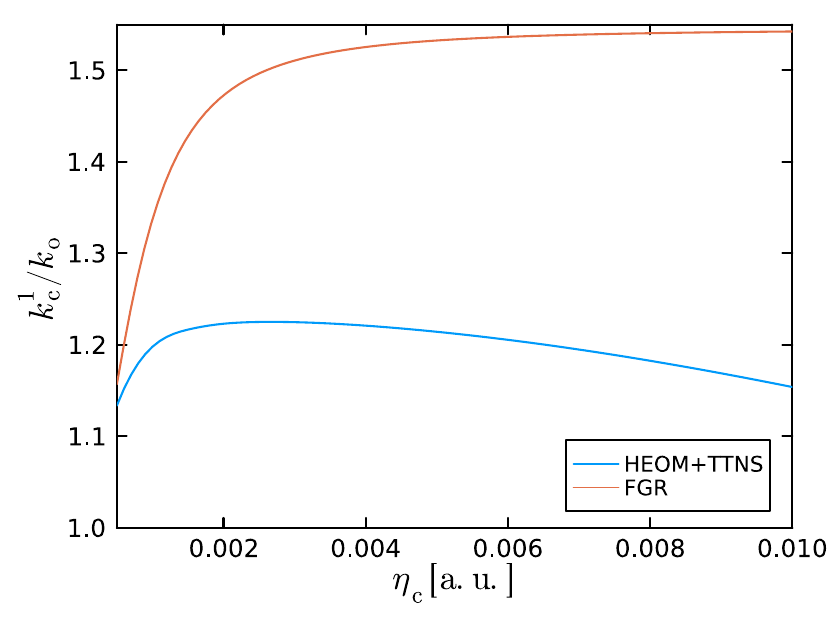}
  \end{minipage}
      \begin{minipage}[c]{0.45\textwidth} 
        \raggedright b) $\lambda_{\rm c}=100\, \mathrm{cm}^{-1}$
    \includegraphics[width=\textwidth]{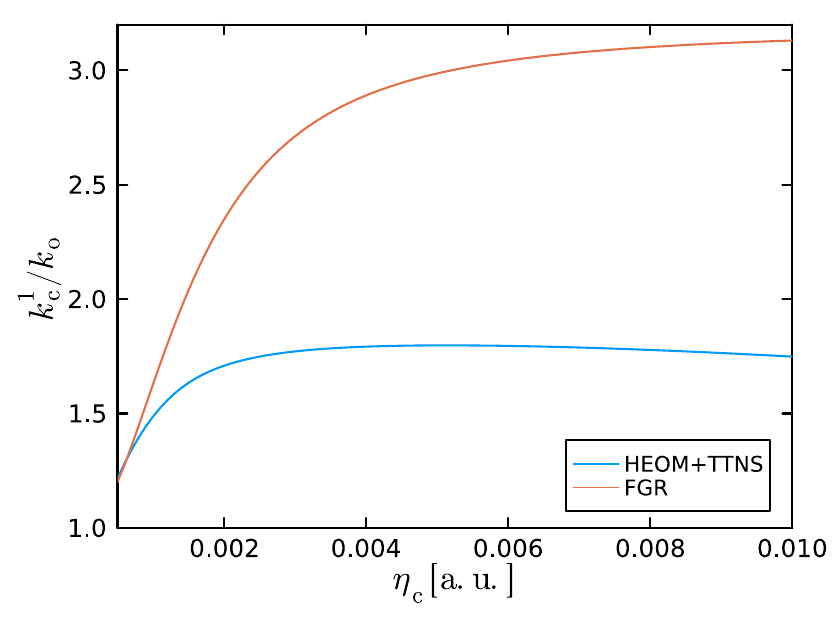}
  \end{minipage}
  \caption{Ratios of reaction rates $k_{\rm c}^1/k_{\rm o}$ in the single-mode limit as a function of the light-matter coupling strength $\eta_{\rm c}$ ranging from 0.0005 to 0.01 a.u. for two different $\lambda_{\rm c}$ in the weak-loss regime. The cavity frequency $\omega_{\rm c}=1185\,\mathrm{cm}^{-1}$ is used. The blue lines correspond to the exact results obtained from the HEOM+TTNS method, and the orange lines are obtained with the modified FGR theory (\Eq{modified_FGR_rate}).} \label{FGR_etac}
\end{figure}

As shown in \Fig{fig2:HEOM_FGR} c1), this modified FGR rate theory corrects the large overestimation of the reaction rates in the single-mode limit, and well captures the peak red-shifting with increasing $\eta_{\rm  c}$. However, in the strong $\eta_{\rm c}$ regime, the modified FGR rates still deviate from the exact results shown in \Fig{fig2:HEOM_FGR} a1). Specifically, for a fixed value of $\lambda_{\rm c}$, the modified FGR rate theory in \Eq{modified_FGR_rate} predicts that the rate constant first scales quadratically with $\eta_{\rm c}$ and then saturates, as illustrated in \Fig{FGR_etac}. In contrast, the exact results show that the reaction rates do not saturate but instead reach a maximum and then gradually decrease with increasing $\eta_{\rm c}$ (the blue lines in \Fig{FGR_etac}). Furthermore, the location of this turnover shifts to larger  $\eta_{\rm c}$ values as $\lambda_{\rm c}$ increases. 

The physical origin of this non-monotonicity lies in the interplay of energy exchange processes among the molecular vibration, the cavity mode, the cavity bath, and the solvent bath. Cavity-induced reaction rate modification mechanism involves the relay of thermal fluctuations: energy from one bath (or an area of the bath) is transferred subsequently via the cavity field and the molecular vibration to another bath (or another area of the bath). When $\lambda_{\rm c}$ is too small, the energy exchange between the cavity and its bath becomes inefficient, restricting the overall reaction rate. Conversely, for a fixed $\eta_{\rm c}$, an overly large $\lambda_{\rm c}$ confines the energy between the cavity and its bath, again limiting the energy transfer to the molecule. This explains the turnover with respect to $\lambda_{\rm c}$. A parallel turnover occurs when $\eta_{\rm c}$ is varied at fixed $\lambda_{\rm c}$. For very small $\eta_{\rm c}$, the molecule–cavity exchange is too weak to mediate efficient transfer; for very large $\eta_{\rm c}$, energy becomes trapped in rapid oscillations between molecule and cavity, reducing the net transfer to the baths. Only at intermediate $\eta_{\rm c}$ is the energy flow balanced across all subsystems, yielding optimal rate enhancement. These results suggest that optimal cavity-modified reactivity requires $\eta_{\rm c}$ and $\lambda_{\rm c}$ to be tuned in a concerted fashion so that all energy-exchange steps among involved components occur on comparable timescales.

Extending to multi-mode cavities, \Fig{fig2:HEOM_FGR} a2), b2), and c2) show the rate modification profile as a function of the free spectral range $\Delta$, obtained with HEOM+TTNS, the standard FGR rate [\Eq{FGR_rate}], and the modified FGR rate [\Eq{modified_FGR_rate}], respectively. The limitations of FGR-type theories become evident here: both the standard and modified FGR frameworks fail completely, missing essential features such as the peak splitting of the rates with respect to the FSR in the near-resonant regime for the neighboring mode. This failure arises because, in the existing FGR framework, the cavity modes are effectively treated as part of the bath, thereby neglecting explicit light–matter hybridization—the very mechanism that produces splitting in the multimode strong-coupling regime. These results demonstrate that the current FGR theory remains inadequate for strong light-matter coupling and multi-mode situations. Our findings call for future theoretical efforts to develop more advanced rate theories applicable to broader parameter regimes and in multi-mode scenarios (at least for Model I), which explicitly incorporate polariton formation, mode–mode interference, and non-perturbative light–matter dynamics, to faithfully describe cavity-modified chemical reactivity. 

\section{Broadening effect from the cavity bath}
\begin{figure}
\centering
    \begin{minipage}[c]{0.45\textwidth}
    \raggedright a) $\lambda_{\rm c}=200\, \mathrm{cm}^{-1}$
    \includegraphics[width=\textwidth]{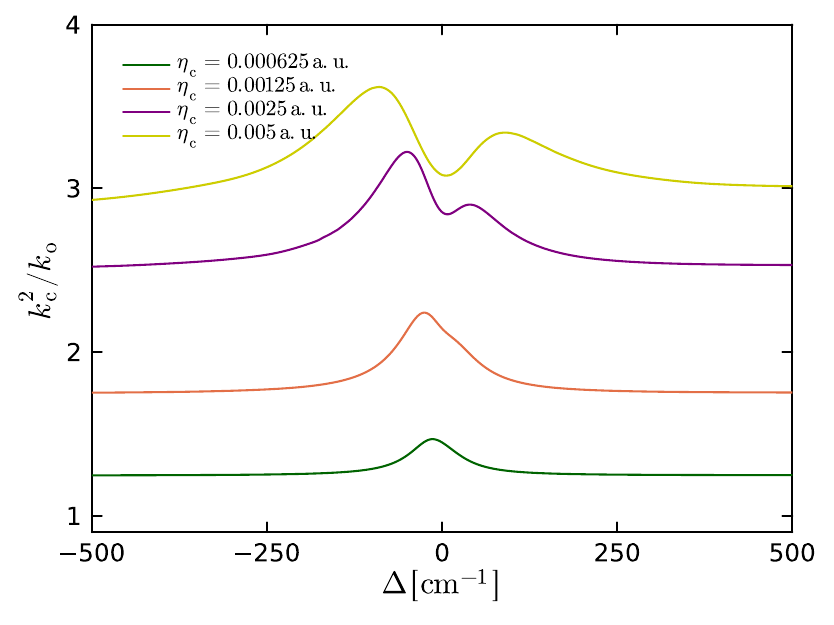}
  \end{minipage}
    \begin{minipage}[c]{0.45\textwidth}
        \raggedright b) $\lambda_{\rm c}=300\,\mathrm{cm}^{-1}$ 
    \includegraphics[width=\textwidth]{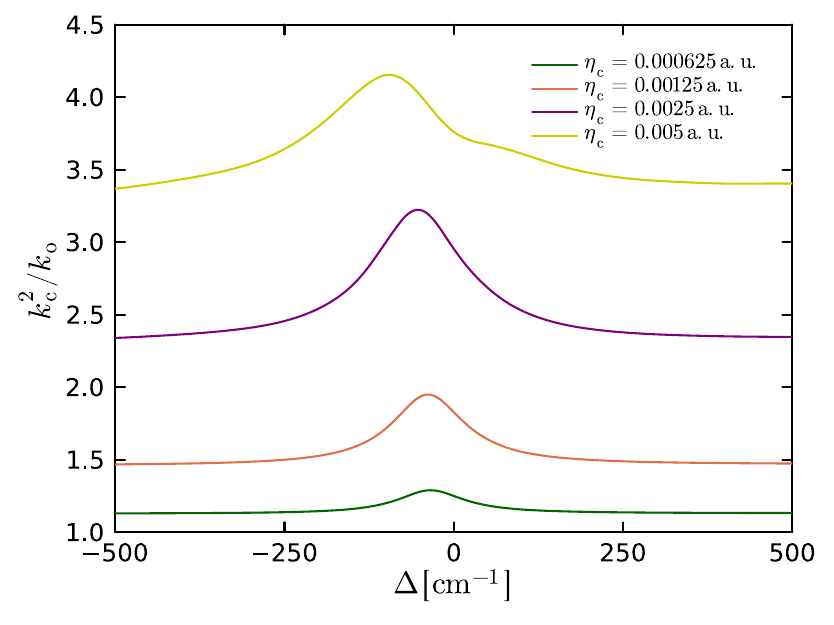}
  \end{minipage}
\caption{Rate modification profile ($k_{\rm c}^2/k_{\rm o}$) for Model I in a two-mode cavity as a function of the free spectral range $\Delta$.   Here, we assume the coupling strengths of two cavity modes to the molecule are the same: $\eta_1=\eta_2=\eta_c$. The results are displayed for different values of $\eta_{\rm c}$ and two different $\lambda_{\rm c}$. The central cavity frequency is fixed at $\omega_{\rm c}=1185\,\mathrm{cm}^{-1}$.} \label{fig3:broadening}
\end{figure}
In the main text, and particularly in Fig. 10~b), we demonstrated the crucial role of the cavity’s coupling strength to its environment, $\lambda_{\rm c}$, in shaping the reaction dynamics. The dependence exhibits a characteristic turnover behavior: as $\lambda_{\rm c}$ increases, the reaction rate first grows, reaches a maximum at an intermediate value, and then decreases. This behavior highlights the importance of energy exchange between the cavity mode and its surrounding bath as an integral component of the cavity-induced reaction pathway, as illustrated in Fig. 5~b). When $\lambda_{\rm c}$ is too weak, this exchange becomes the rate-limiting step, creating a bottleneck for the overall reaction dynamics. At intermediate $\lambda_{\rm c}$, the exchange is optimally balanced, allowing the cavity to mediate energy transfer most efficiently and thereby maximize the rate enhancement. However, when $\lambda_{\rm c}$ is too strong, excessive photon leakage from the cavity disrupts coherent light–matter interactions, thereby reducing the rate.  Importantly, this turnover behavior arises in both the single-mode and two-mode scenarios, indicating that it is a robust feature of cavity-modified chemistry.

In addition, increasing $\lambda_{\rm c}$ also induces spectral broadening of the polaritonic resonances. This effect is illustrated in \Fig{fig3:broadening}, which shows the reaction rate modification profile in a two-mode cavity as a function of the FSR $\Delta$ for two larger values of $\lambda_{\rm c}$. In a weak-loss cavity, such as when $\lambda_{\rm c}=100,\mathrm{cm}^{-1}$, distinct peak splittings appear in the rate modification profile (see \Fig{fig2:HEOM_FGR}~a2)), reflecting the formation of distinct hybridized light–matter states. However, when $\lambda_{\rm c}$ becomes sufficiently large for a given light–matter coupling strength, these sharp features are washed out. Strong dissipation broadens the polaritonic states to the point that they overlap significantly and effectively merge into a degenerate manifold. In this regime, the fine structure associated with hybridization through one cavity mode is no longer discernible by the other mode, and the peak splitting that characterizes the multi-mode strong coupling regime vanishes.

\section{Independent vs. shared cavity bath}
\begin{figure}
\centering
    \begin{minipage}[c]{0.45\textwidth} 
    \includegraphics[width=\textwidth]{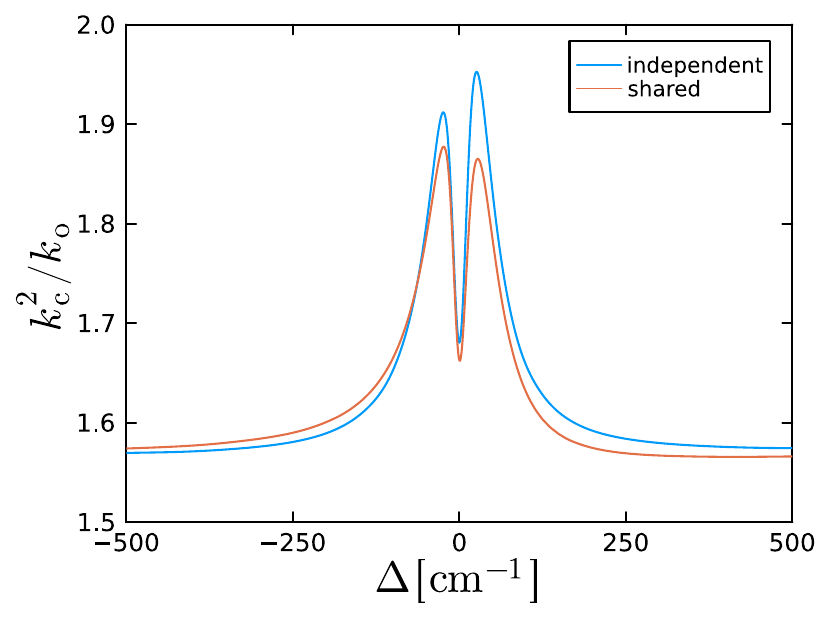}
  \end{minipage}
\caption{Rate modification profile ($k_{\rm c}^2/k_{\rm o}$) for Model I in a two-mode cavity as a function of the free spectral range $\Delta$ for two different scenarios. The blue line corresponds to the case where each cavity mode is coupled to its own independent bath, whereas the orange line corresponds to the case where both cavity modes are coupled to a single shared cavity bath.  Here, we assume the coupling strengths of two cavity modes to the molecule are the same: $\eta_1=\eta_2=0.00125$\au Besides, the central cavity frequency is fixed at $\omega_{\rm c}=1185\,\mathrm{cm}^{-1}$ and $\lambda_{\rm c}=100\,\mathrm{cm}^{-1}$.} \label{fig4:sharedbath}
\end{figure}
In the main text, we assumed that each cavity mode is coupled to its own independent bath. However, the lossy environment need not be strictly local. In practice, it is possible that the same dissipative environment simultaneously couples to multiple cavity modes. To investigate this scenario, we consider the case where all cavity modes are coupled to a single shared cavity bath. Specifically, Eq. (3) in the main text is adapted as
\begin{equation}
\label{environmentHamiltonian}
H_{\rm E}^{\rm c} = \sum_{k} \frac{P_{{\rm c}k}^2}{2}+ \frac{1}{2}\omega_{{\rm c}k}^2 \left(Q_{{\rm c}k}+\sum_i\frac{g_{ik}}{\omega_{{\rm c}k}^2}x_{i}\right)^2 
=  \sum_{k} \frac{P_{{\rm c}k}^2}{2}+ \frac{1}{2}\omega_{{\rm c}k}^2 Q^2_{{\rm c}k} +\sum_i g_{ik}Q_{{\rm c}k}x_{i}  +\sum_i \frac{g^2_{ik}}{\omega_{{\rm c}k}^2} x_{i}^2 
 +\frac{1}{2\omega_{{\rm c}k}^2}\sum_{i\neq j}g_{ik}g_{jk}x_{i} x_{j}. 
\end{equation}

In this setting, the shared bath introduces cross-interaction between different cavity modes (last term in Eq. \ref{environmentHamiltonian}). This means that the cavity modes are not only indirectly coupled through their mutual interaction with the molecular vibration, but also through their shared coupling to the common bath. To evaluate the impact of this additional indirect mode–mode interaction, we compare the rate modification profiles $k_{\rm c}^2/k_{\rm o}$ as a function of the FSR $\Delta$ for the two scenarios: independent cavity baths versus a shared cavity bath. The results are shown in \Fig{fig4:sharedbath}.

We find that collective coupling of the cavity modes to a shared bath slightly enhances the reaction rate when $\Delta \ll 0$, but slightly suppresses the rate enhancement when the neighboring cavity mode frequency is close to or larger than the central cavity frequency. Nevertheless, the key qualitative features remain robust. The characteristic peak splitting and the associated splitting gap persist, indicating that the essential physics of multi-mode cavity-enhanced chemistry is not qualitatively altered by whether the cavity baths are independent or shared.

\section{Suppressed rate enhancement upon the inclusion of additional cavity mode}
In most of the cases and parameter regimes we have examined, the inclusion of a second cavity mode leads to an overall enhancement of the reaction rate. This suggests, {\it post hoc}, that the mechanism depicted in Fig. 5--where the cavity introduces an additional reaction pathway that facilitates the transfer of energy fluctuation from one thermal bath via the cavity mode and molecule to the other thermal bath--is typically dominant. However, while the prevailing trend is rate enhancement upon adding a second cavity mode, there do exist regimes--particularly under strong light–matter coupling in the weak cavity damping limit--where the opposite occurs, i.e., the reaction rate is reduced by the presence of the additional cavity mode. Two examples are discussed below.

\begin{figure}
\centering
    \begin{minipage}[c]{0.45\textwidth} 
      \raggedright a) $\lambda_{\rm c}=50\, \mathrm{cm}^{-1}$
    \includegraphics[width=\textwidth]{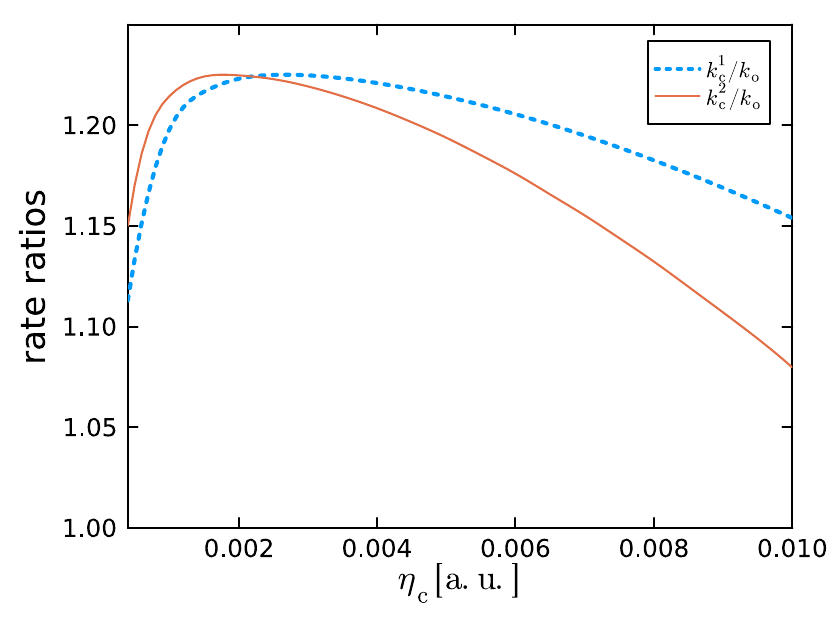}
  \end{minipage}
      \begin{minipage}[c]{0.45\textwidth} 
        \raggedright b) $\lambda_{\rm c}=100\, \mathrm{cm}^{-1}$
    \includegraphics[width=\textwidth]{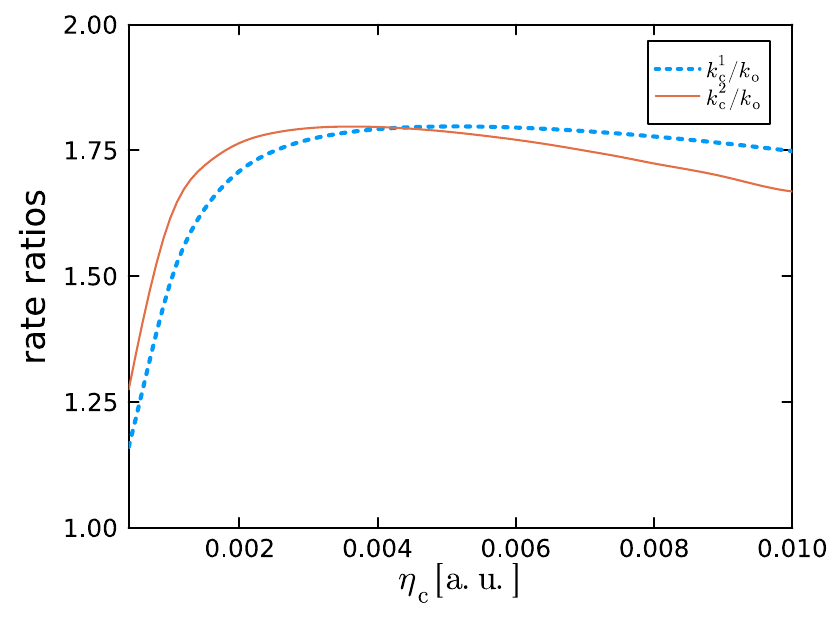}
  \end{minipage}
  \caption{Ratios of reaction rates $k_{\rm c}^{1/2}/k_{\rm o}$ inside and outside the cavity for Model I as a function of the light-matter coupling strength $\eta_{\rm c}$ (ranging from 0.0005 to 0.01\au). The blue dotted line corresponds to the single-mode case $k_{\rm c}^{1}/k_{\rm o}$ and the cavity frequency is fixed at $\omega_{\rm c}=1185\,\mathrm{cm}^{-1}$. The orange solid line corresponds to the two-mode case with degenerate modes: $\omega_{1}=\omega_{2}=\omega_{\rm c}=1185\,\mathrm{cm}^{-1}$ and equal coupling strength $\eta_{1}=\eta_2=\eta_{\rm c}$. Two different cavity loss strengths are used: $\lambda_{\rm c}=50\,\mathrm{cm}^{-1}$ in a) and $\lambda_{\rm c}=100\,\mathrm{cm}^{-1}$ in b).} \label{fig5:twomodes}
\end{figure}

\begin{figure}
\centering
    \begin{minipage}[c]{0.45\textwidth} 
    \includegraphics[width=\textwidth]{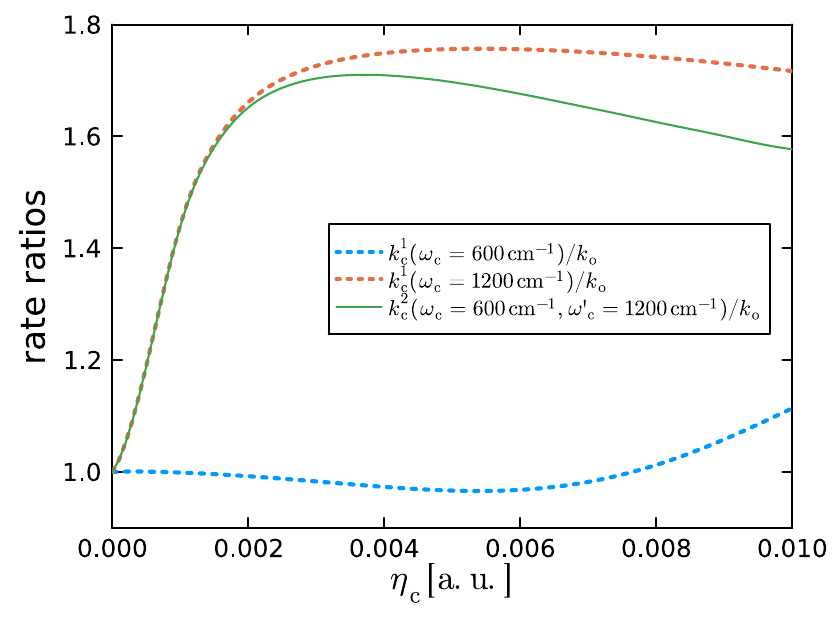}
  \end{minipage}
  \caption{Ratios of reaction rates $k_{\rm c}^{1/2}/k_{\rm o}$ inside and outside the cavity for Model I as a function of the light-matter coupling strength $\eta_{\rm c}$. The blue and orange dotted line correspond to the single-mode case $k_{\rm c}^{1}/k_{\rm o}$ for the cavity frequencies fixed at $\omega_{\rm c}=600\,\mathrm{cm}^{-1}$ and $\omega_{\rm c}=1200\,\mathrm{cm}^{-1}$, respectively. The green solid line corresponds to the two-mode case where the frequency of the two modes are $\omega_{\rm c}=600\,\mathrm{cm}^{-1}$  and $\omega'_{\rm c}=1200\,\mathrm{cm}^{-1}$ with equal light-matter coupling strengths $\eta_{1}=\eta_2=\eta_{\rm c}$. The cavity loss strength  $\lambda_{\rm c}=100\,\mathrm{cm}^{-1}$ is used.  } \label{fig6:twodifferentmodes}
\end{figure}

First, consider the mechanisms illustrated in Fig. 5 and Fig. 9 of the main text, where photon exchange between the cavity mode and the molecular vibration is, among others, a crucial step in the cavity-mediated reaction pathway. For a fixed $\lambda_{\rm c}$, which settles the energy exchange rate between the cavity mode and its bath, increasing the light–matter coupling strength $\eta_{\rm c}$ also produces a turnover in the reaction rate, as shown in \Fig{fig6:twodifferentmodes}. The turning point is set by the balance between $\eta_{\rm c}$ and $\lambda_{\rm c}$. The maximum rate occurs when all relevant energy exchange processes along the cavity-induced pathway occur on comparable timescales, allowing the whole system to operate in a concerted and efficient manner. When $\eta_{\rm c}$ is too small, the probability of relaying energy from the cavity to the molecule is low, resulting in only a weak rate enhancement. Conversely, if $\eta_{\rm c}$ is too large, energy becomes trapped in rapid back-and-forth exchanges between the cavity and the molecule, which hinders progress along the entire energy transfer pathway and reduces the overall rate. Under such conditions, the inclusion of a second cavity mode exacerbates this trapping effect, further reducing the reaction rate, as illustrated in \Fig{fig5:twomodes}.

A second situation arises when two cavity modes are both nearly resonant to a vibrational transition, and the lower-frequency mode is supported via an effective two-photon process. In the very strong light–matter coupling regime, the addition of this low-frequency mode can actually suppress the reaction rate, as shown in \Fig{fig6:twodifferentmodes}. Interestingly, even though each of these modes individually enhances the rate when considered in isolation (single-mode limit), their combined effect can be destructive in the multi-mode case. This highlights the subtle role of mode–mode interplay in determining the overall reactivity.

In summary, while multi-mode coupling most often enhances reaction rates in our study, there do exist parameter regimes where the opposite trend emerges. These findings underscore the rich and nontrivial dynamics that arise beyond the single-mode picture, where cooperative, competitive, and interference effects among multiple cavity modes collectively shape the outcome of cavity-modified chemistry.

\end{document}